\documentclass[a4paper,twocolumn,11pt,accepted=2026-04-07]{quantumarticle}
\pdfoutput=1
\usepackage[utf8]{inputenc}
\usepackage[english]{babel}
\usepackage[T1]{fontenc}
\usepackage{amsmath}

\usepackage[]{qcircuit}
\usepackage{multirow}
\usepackage{booktabs}

\usepackage{amsmath,amssymb,amsthm,bm,amsfonts,mathrsfs,bbm}

\usepackage{dsfont}

\usepackage{graphicx}
\usepackage{float}
\usepackage{tikz}
\usepackage[export]{adjustbox}

\usepackage{float}
\usepackage{algorithm}
\usepackage[noend]{algpseudocode}

\usepackage{bbold}
\usepackage{braket}
\usepackage{etoolbox}
\usepackage{hyperref}
\usepackage{cleveref}

\usepackage{placeins}

\makeatletter
\providecommand\@afterenddocumenthook{}
\makeatother
\begin{document}

\title{Swap Network Augmented Ansätze on Arbitrary Connectivity}

\author{Teodor Parella-Dilmé}
\affiliation{ICFO - Institut de Ciencies Fotoniques, The Barcelona Institute of Science and Technology, Av. Carl Friedrich Gauss 3, 08860 Castelldefels (Barcelona), Spain}
\orcid{0009-0000-7593-2417}
\thanks{\texttt{teodor.parella@icfo.eu}}

\author{Jakob~S.~Kottmann}
\affiliation{Institute for Computer Science, University of Augsburg, Germany}
\affiliation{Center for Advanced Analytics and Predictive Sciences, University of Augsburg, Germany}
\orcid{0000-0002-4156-2048}
\thanks{\texttt{jakob.kottmann@uni-a.de}}

\author{Antonio Acín}
\affiliation{ICFO - Institut de Ciencies Fotoniques, The Barcelona Institute of Science and Technology, Av. Carl Friedrich Gauss 3, 08860 Castelldefels (Barcelona), Spain}
\affiliation{ICREA-Institucio Catalana de Recerca i Estudis Avancats, Lluis Companys 23, 08010 Barcelona, Spain}
\orcid{0000-0002-1355-3435}
\thanks{\texttt{antonio.acin@icfo.eu}}
\maketitle

\begin{abstract}
  Efficient parametrizations of quantum states are essential for trainable hybrid classical-quantum algorithms. A key challenge in their design consists in adapting to the available qubit connectivity of the quantum processor, which limits the capacity to generate correlations between distant qubits in a resource-efficient and trainable manner. In this work we first introduce an algorithm that optimizes qubit routing for arbitrary connectivity graphs, resulting in a swap network that enables direct interactions between any pair of qubits. We then propose a co-design of circuit layers and qubit routing by embedding the derived swap networks within layered, connectivity-aware ansätze. This construction significantly improves the trainability of the ansatz, leading to enhanced performance with reduced resources. We showcase these improvements through ground-state simulations of strongly correlated systems, including spin-glass and molecular electronic structure models. Across exemplified connectivities, the swap-enhanced ansatz consistently achieves lower energy errors using fewer entangling gates, shallower circuits, and fewer parameters than standard layered-structured baselines. Our results indicate that swap network augmented ansätze provide enhanced trainability and resource-efficient design to capture complex correlations on devices with constrained qubit connectivity.
\end{abstract}

\section{Introduction}

\begin{figure*} [t]                    
\centering
\includegraphics[width=2\columnwidth, angle=0]{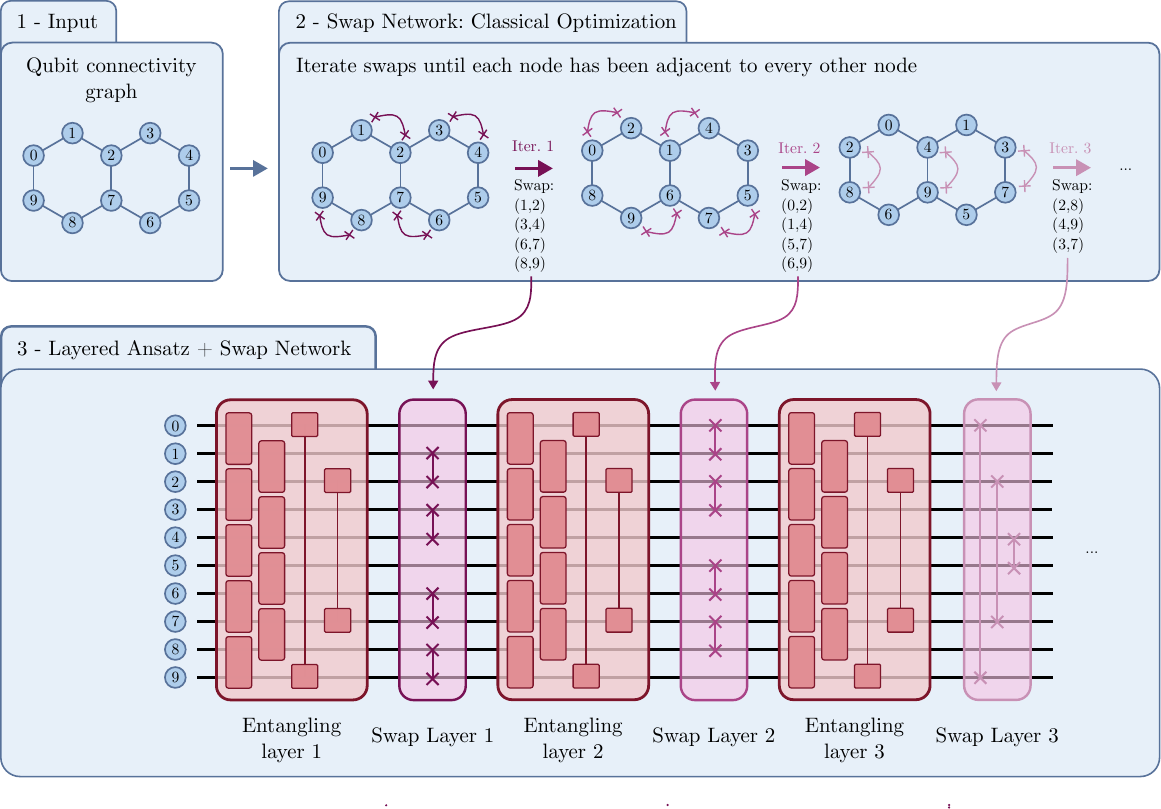}
\caption{Diagram of the proposed algorithm for designing a layered ansatz embedded within a swap network with optimized qubit routing in arbitrary connectivity. \textbf{1)} The available qubit connectivity graph is taken as input. \textbf{2)} A classical optimization algorithm designs an optimized swap network for the given connectivity constraints. The optimized sequence of swaps ensures that by the end of the process each node has been adjacent to every other node at least once. \textbf{3)} The layered ansatz is embedded within the swap network, enabling direct information exchange between all qubits on the constrained connectivity hardware.
}
\label{fig:Infographic}
\end{figure*} 
In variational quantum algorithms (VQAs) a parametrized quantum circuit (PQC) is optimized through a classical routine to solve optimization tasks \cite{Cerezo_Arrasmith_Babbush_Benjamin_Endo_Fujii_McClean_Mitarai_Yuan_Cincio, McClean_Romero_Babbush_Aspuru-Guzik_2016}. The structure and initialization of an ad-hoc PQC play a crucial role in the optimization and overall algorithm performance, as they influence both the accessible state space and the optimization landscape \cite{Tilly_Chen_Cao_Picozzi_Setia_Li_Grant_Wossnig_Rungger_Booth, Sim_Johnson_Aspuru‐Guzik_2019}. Effective ansatz design is best guided by the problem’s structure (when available) and the limitations of the available hardware. For instance, physically-inspired ansätze exploit known symmetries to reduce the solution space \cite{romero2018strategiesquantumcomputingmolecular}, while hardware-efficient ansätze (HEA) leverage native gates and direct qubit connectivity to minimize resource requirements \cite{Kandala_2017}.
\\
\\
Regarding hardware limitations, constrained qubit connectivity represents a major challenge: two-qubit gates can only be applied between neighbouring qubits in the device’s connectivity. A common practice is to insert SWAP gates via routing strategies \cite{li2019tacklingqubitmappingproblem,https://doi.org/10.4230/lipics.tqc.2019.5,murali2019noiseadaptivecompilermappingsnoisy,zhu2025quantumcompilerdesignqubit} to bring qubits together whenever a two-qubit gate needs to be applied. However each SWAP (implemented in practice via multiple entangling gates) adds an overhead in noise and circuit depth \cite{Holmes_Johri_Guerreschi_Clarke_Matsuura_2020}. In contrast, the aforementioned HEA are another workaround to the constrained connectivity problem by using ansätze that only entangle adjacent qubits in the connectivity using native gates. While this avoids SWAP gates, it severely limits the spread of correlations across the processor, therefore compromising trainability when the target state has long-range entanglement \cite{McClean_2018}. 
\\
\\
An alternative strategy to overcome connectivity limitations is the use of swap networks in a linear qubit connectivity \cite{PhysRevLett.120.110501, ogorman2019generalizedswapnetworksnearterm, Anselmetti_Wierichs_Gogolin_Parrish_2021}. These are predetermined sequence of SWAP gates arranged in layers, designed so that after the sequence every qubit has effectively moved next to every other qubit at least once, enabling all-to-all connectivity. The network systematically permutes the qubits across the device so that any logical pair can eventually interact, without needing on-the-fly routing decisions. The fixed swap schedule deterministically guarantees all desired interactions occur, offering potential applicability in unstructured or repetitive circuits, such as Trotterized dynamics \cite{hagge2022optimalfermionicswapnetworks} or alternating-layer algorithms like the Quantum Approximate Optimization Algorithm (QAOA)~\cite{PhysRevResearch.4.033028,PhysRevResearch.6.043279}.
\\
\\
In this work, we first introduce an algorithm for generating optimized swap networks for arbitrary qubit connectivity, which is available at
\cite{AutoSwap}. We then propose integrating the optimized swap networks with hardware-adapted layer-structured ansätze made of repeated blocks of local entanglers [Fig.\ref{fig:Infographic}]. The net effect is a hardware-aware ansatz that effectively enables all-to-all connectivity with reduced SWAP overhead compared to dynamical compilers such as SABRE \cite{li2019tacklingqubitmappingproblem, zou2024lightsabrelightweightenhancedsabre}. Our procedure avoids the use of measurement-intensive Adaptive
Derivative-Assembled Pseudo-Trotter ansatz (ADAPT) methods \cite{Grimsley_Economou_Barnes_Mayhall_2019,PRXQuantum.2.020310,Yordanov_Armaos_Barnes_Arvidsson-Shukur_2021}, therefore avoiding metaoptimization during the ansatz construction. 
\\
\\
We demonstrate through variational ground-state simulations of highly correlated systems that the swap network enhanced ansatz significantly improves both trainability and performance, all while maintaining low resource overhead. These gains are showcased across benchmark problems, including quantum spin-glass Hamiltonians and the electronic structure of the strongly correlated p-benzyne biradical. For a fixed resource budget of gate count, circuit depth, or number of parameters, the swap network enhanced ansatz consistently achieves lower energy errors and improved convergence compared to the HEA.This translates into higher noise resilience: at the same circuit depth or entangling gate count, and therefore comparable accumulated noise, it achieves lower errors and more reliable convergence. Across various connectivity graphs, we find that long-range correlations are captured more efficiently, underscoring the advantage of incorporating optimized swap networks in the ansatz design.

\section{Preliminaries}

\subsection{Circuit Compilation} 
Circuit compilation is the process of transforming an abstract quantum circuit, defined as an ordered sequence of ideal gates,
\[
U(\boldsymbol{\theta}) = \prod_{k=1}^{K} u_k\!\bigl(\theta_{k}\bigr),  \qquad \boldsymbol{\theta} \in \mathbb{R}^K,
\]
into a hardware compatible circuit $U'(\boldsymbol{\theta})$ that can be executed on a specific quantum processor. Here $u_k$ denotes the $k$th circuit gate, which may be a one- or two-qubit operation, and may be either fixed or parametrized. This transformation is necessary because quantum hardware supports only a limited set of native gates and allows two-qubit operations only between certain pairs of connected qubits. The compilation process typically comprises two main tasks  (may include additional ones):

\begin{enumerate}
    \item \textbf{Gate decomposition} translates or approximates all high-level operations into sequences of native gates, ensuring that every operation conforms to the hardware’s supported gate set.

    \item \textbf{Qubit routing} adapts the logical circuit to the physical qubit layout by initially mapping logical qubits onto physical qubits and then routing them by inserting swap gates where necessary, so that two-qubit operations occur only between adjacent qubits in the connectivity.
\end{enumerate}
Here, we consider gates in the abstract circuit $U$ to be native, and take a closer look at the qubit routing step. Although essential, qubit routing imposes a significant burden on the resulting circuit by introducing numerous swap gates. These gates decompose into multiple error-prone two-qubit operations within the native gate set (Appendix \ref{AppendixB}). The additional two-qubit operations not only increase circuit depth but also represent a primary source of noise, ultimately degrading computational fidelity on near-term devices. Therefore, minimizing swap overhead is essential for enhancing the efficiency of quantum simulations, reducing both noise and circuit depth to improve overall performance.

\subsection{Hardware-Efficient Ansätze}
To mitigate the detrimental effects of extensive swap overhead, a promising strategy is to design circuits that align with the hardware’s native constraints. HEA embody this approach by directly using the available native gate set and restricting two-qubit operations to adjacent qubits in the hardware connectivity \cite{Kandala_2017}. The HEA is typically implemented in an alternating-layer construction, where layers of gates are arranged in a structured, repeating pattern. By doing so, HEA eliminate the need for circuit compilation and the associated swap overhead, preventing additional swaps that would otherwise increase circuit depth and noise. 
\\
\\
This improvement, however, comes at the cost of sacrificing all-to-all connectivity. Since two-qubit gates can only be applied to adjacent qubits, establishing correlations across distant regions of the processor requires a chain of intermediate interactions that relay information in the causal cone. Capturing long-range correlations in such architectures is particularly challenging because many parameters must simultaneously align to effectively correlate distant qubits, an effect that can lead to local minima or barren plateaus in the HEA cost landscape \cite{McClean_2018,Larocca_Thanasilp_Wang_Sharma_Biamonte_Coles_Cincio_McClean_Holmes_Cerezo_2025}. This intricate interdependence complicates the optimization and can severely limit ansatz trainability in the presence of long-range correlations \cite{Holmes_Johri_Guerreschi_Clarke_Matsuura_2020}. In particular, Ref.\cite{Leone2024practicalusefulness} establishes that the HEA can remain trainable when the input states obey an area law of entanglement; conversely, for inputs with long-range correlations satisfying a volume law, the loss values concentrate exponentially around a trivial value, rendering the HEA effectively untrainable and inexpressible.
\\
\\
Despite these challenges, tailoring ansatz design to the underlying hardware remains a crucial strategy for minimizing circuit depth and mitigating noise, thereby enabling practical ansätze for quantum simulation.

\subsection{Swap Networks} 
An alternative resource efficient approach that preserves effective all to all connectivity while minimizing compilation induced swap overhead is the use of swap networks~\cite{PhysRevLett.120.110501, ogorman2019generalizedswapnetworksnearterm, Anselmetti_Wierichs_Gogolin_Parrish_2021}. Swap networks are predetermined sequences of swap gates that systematically reconfigure the qubits throughout the circuit execution. Unlike ad hoc routing strategies that insert swaps on a case-by-case basis in the compilation step whenever non-adjacent qubits need to interact, a swap network adheres to a fixed routing schedule that guarantees every required qubit interaction will eventually occur. This instance-agnostic strategy is especially valuable when the circuit exhibits an unstructured sequence of gates, as is common in Trotterized simulations \cite{hagge2022optimalfermionicswapnetworks} or the alternating layers of QAOA~\cite{ogorman2019generalizedswapnetworksnearterm}. By optimally layering swap operations, swap networks completely avoid the circuit compilation step by providing a robust, hardware-adapted method to achieve the desired entanglement among qubits with minimal swap overhead.
\\
\\
In the following, we optimize the design of swap network protocols for arbitrary hardware graphs, enabling effective all-to-all connectivity ansätze on modern processors with any connectivity.

\section{Swap Network Optimization For Arbitrary Qubit Connectivity}

\subsection{Overview}
A diagram illustrating the proposed algorithm, featuring the construction of a hardware-friendly layered ansatz embedded in an optimized swap network, is shown in Fig.\ref{fig:Infographic}. The algorithm takes the qubit connectivity graph as input. A classical optimization process determines how to construct a circuit with a reduced number of swap layers that ensure that every qubit pair is connected at least once during the routing. The resulting swap layers from the classical optimization define the swap network, which is then interwoven between the layers of the hardware-friendly layered ansatz. Ultimately, a swap enhanced ansatz with optimized routing sequence is obtained, enabling direct information exchange between all pairs of qubits in the system. In the following, we detail each subroutine of the algorithm.

\subsection{Qubit Display as a Labelled Graph}

We define a labelled and undirected graph $\mathcal{G} = (V, E, \mathcal{L})$, where:

\begin{itemize}
    \item $V = \{ v_i \}_{i=1}^{n}$ is a set of $n$ labelled vertices.
    \item $E \subseteq \{ (x, y) \in V \times V \mid x \neq y \}$ is a subset of all the pairs of different vertices.
    \item $\mathcal{L}: V \to L$ denotes a bijective labelling function that assigns each vertex $v_i \in V$ a label $\mathcal{L}(v_i) \in L$, with the label set $L = \{ 1, 2, \ldots, n \}$.
\end{itemize}

The set $V$ represents the physical sites of the hardware, while the set $E$ defines the connections between these sites, representing the possible entangling interactions. Furthermore, $\mathcal{L}(v_i)$ denotes the label assigned to the qubit located at each physical site $v_i$.

\subsection{Qubit swapping as Graph Relabelling Functions}
We now introduce a graph operation simulating the behaviour of a swap gate, corresponding to a label-swapping function $\mathcal{S}(\mathcal{G}, e_{i,j}) = \mathcal{G}'$, $e_{i,j}=(v_i,v_j)\in E$ which swaps the labels of the connected vertices $v_i$ and $v_j$ in a graph $\mathcal{G}$. The resulting graph $\mathcal{G}' = (V, E, \mathcal{L}')$ is characterized by the updated labelling:
\begin{itemize}
    \item $\mathcal{L}'(v_i) = \mathcal{L}(v_j)$,
    \item $\mathcal{L}'(v_j) = \mathcal{L}(v_i)$,
    \item $\mathcal{L}'(v_k) = \mathcal{L}(v_k)$ for $k \neq i,j$.
\end{itemize}
To emulate a depth-1 layer of swap gate operations, we define $\mathcal{R}^{1}$ as a layer of swap-relabelling functions of depth 1. This function consists of a concatenation of commuting label-swapping functions $\{\mathcal{S}_i\}_{i=1}^m$, each acting on different vertices. Specifically:
\begin{equation*}
\mathcal{R}^{1} = \mathcal{S}_m \circ \mathcal{S}_{m-1} \circ \ldots \circ \mathcal{S}_2 \circ \mathcal{S}_1,
\end{equation*}
Each $\mathcal{S}_i$ represents a distinct label-swapping operation, ensuring that two different operations do not act on the same vertex. This means each node is either swapped once with a neighbour or remains unchanged (only one swap per node).

Ultimately, a number $k$ of different swap-relabelling layer functions of depth 1, can be concatenated to yield a swap-relabelling multilayer function of depth $k$:
\begin{equation*}
\mathcal{R}^k = \mathcal{R}^1_k \circ \mathcal{R}^1_{k-1} \circ \ldots \circ \mathcal{R}^1_2 \circ \mathcal{R}^1_1.
\end{equation*}
Such a multilayer function simulates an arbitrary swap layer of depth $k$ within the available connectivity. The motivation behind $\mathcal{R}^k$ is that performing a single swap within the connectivity does not fully maximize the potential for creating new interactions before applying an entangling gate to neighbouring qubits. If the swap depth is limited to one, the swapped qubits will remain neighbours. Therefore, a depth of $k \geq 2$ is preferred, as it allows for a greater number of new qubit interactions to emerge.

\begin{figure} [h!]                    
\centering
\includegraphics[width=1\columnwidth, angle=0]{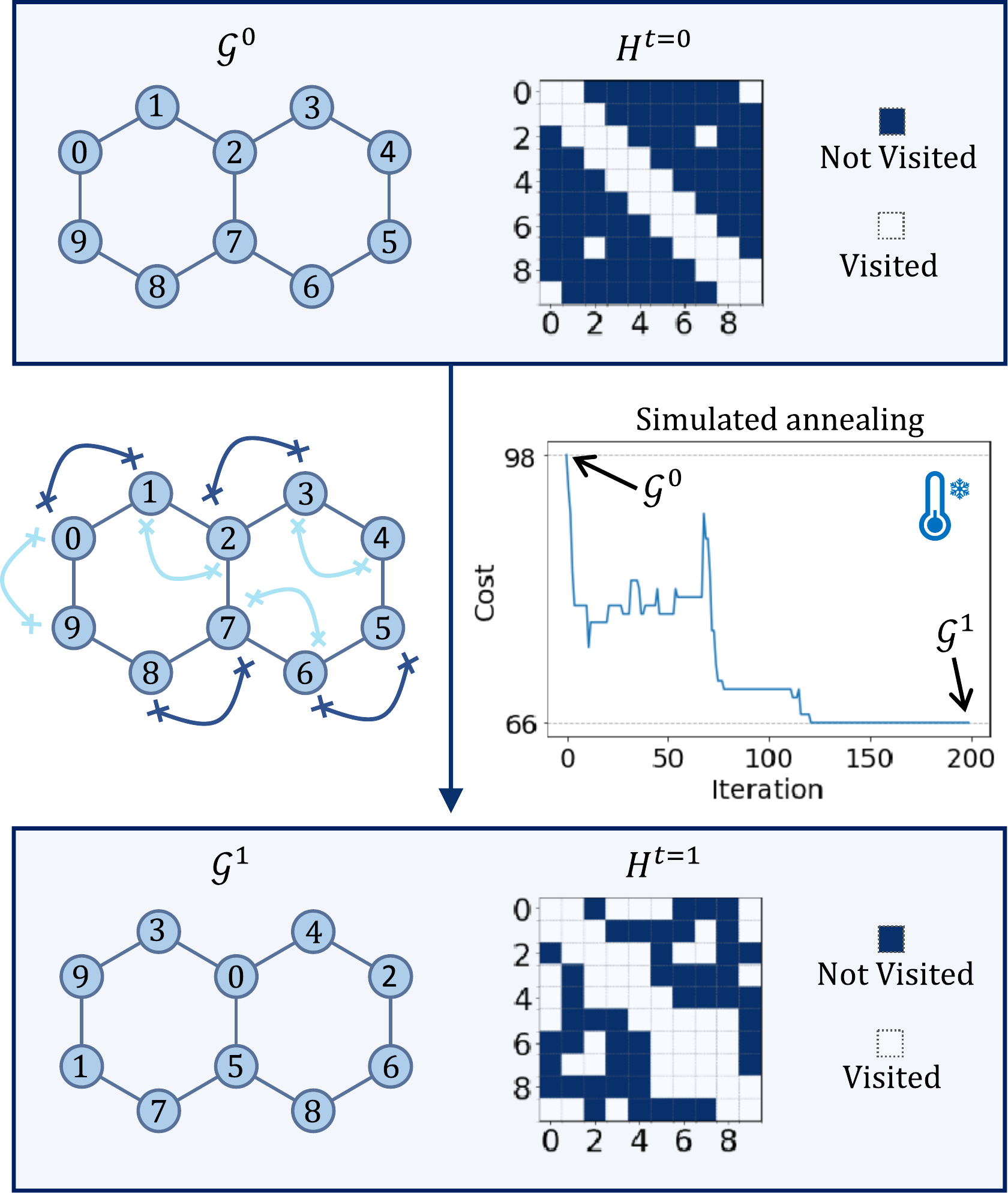}
\caption{Exemplified algorithm step with \(k=2\), that optimizes an initial graph \(\mathcal{G}_0\) into \(\mathcal{G}_1\). The initial history matrix \(H^{t=0}\) equals \(\mathcal{G}_0\) adjacency matrix, where dark blue values (\(H^{t=0}_{i,j} = 1\)) indicate that indices \(i,j\) have yet to become adjacent, and white values (\(H^{t=0}_{i,j} = 0\)) indicate prior adjacency. A simulated annealing optimization modifies the graph by adding or removing swap gates, with up to \(k=2\) swap layers. The cost function minimization is shown at the middle, reducing from \(C(\mathcal{G}_0)=98\) to \(C (\mathcal{G}_1)=66\). The optimized $k=2$ swap sequence morphing \(\mathcal{G}_0\) to \(\mathcal{G}_1\) (left) shows dark swaps in the first layer and light swaps in the second. The final graph \(\mathcal{G}_1\) introduces new direct interactions, and its updated history matrix \(H^{t=1}\) has fewer non visited entries, indicating reduced remaining interactions. To extract the complete swap network, this process can be repeated iteratively until all history matrix entries reach zero.}
\label{fig:SimulatedAnnealing}
\end{figure} 

\subsection{The History Matrix}
A key component of the algorithm is the \textit{history matrix}, denoted as $H$. At each iteration step $t$, the history matrix $H^t$ records the adjacency history of qubit interactions, tracking which pairs of qubits have already been nearest neighbours. Specifically, an entry \( H^t_{i,j} = 0 \) indicates that qubits \( i \) and \( j \) have already shared a direct connection at iteration $\leq t$, while \( H^t_{i,j} = 1 \) indicates they have not yet interacted directly. At the initial step,  \( H^{t=0} \) is complementary to the adjacency matrix of the labelled graph. As the algorithm progresses, successive qubit relabelling allows for new direct connections to form, and the history matrix is updated from \( H^t \) to \( H^{t+1} \). This process continues until all entries are zero, signifying that every qubit pair has been directly connected at least once.

\subsection{Cost Function}
Given the iteration $t$, with associated graph \( \mathcal{G}^t \) and history matrix \( H^t \), we seek to quantify the suitability of a relabelled graph resulting from a swap relabelling function of depth $k$, \( \mathcal{G}^{t+1} = \mathcal{R}^k(\mathcal{G}^t) \) using a cost function \( \mathcal{C} \). The cost function should satisfy the two following objectives:
\begin{enumerate}
    \item Maximize the number of new direct connections between labels.
    \item Minimize the distance between labels that have yet to connect in forthcoming steps.
\end{enumerate}

Let $d_{ij}$ denote the shortest path distance between labels $i$ and $j$ in the graph $\mathcal{G}^t$. We can then define the following cost function on the graph $\mathcal{G}^t$ with history matrix $H^t$:
\begin{equation}
    \mathcal{C}(\mathcal{G}^t,H^t)=\sum_{i,j}H^t_{ij}d_{ij}^{\alpha},
    \label{CostFunction}
\end{equation}
where the $\alpha$ exponent is an hyperparameter. The minimization of this cost function addresses the two key requirements. Firstly, it maximizes the number of new direct connections by summing the terms \( H_{ij} d_{ij}^{\alpha} \), which favours configurations where \( H_{ij} \) values shift from 1 to 0. This aspect of the function encourages the minimization of qubit pairs that have yet to interact. Secondly, the term \( d_{ij}^{\alpha} \) penalizes qubit pairs that remain unconnected and are positioned far apart, thereby promoting closer arrangements within the connectivity graph to facilitate direct connections in subsequent steps.
\\
\\
Note that $\alpha$ is a tunable hyperparameter. Larger values (e.g., $\alpha = 2$) favor clustering distant qubits together rather than introducing new interactions, whereas smaller values (e.g., $\alpha = 1/2$) promote the creation of new interactions. The value of $\alpha$ influences the convergence of the algorithm, and its optimal choice depends on the given qubit connectivity graph. Note that $\alpha$, and therefore the cost function, can be changed during the optimization. Small values help early by promoting local interactions, while larger values help later by bringing distant qubits together and speeding up convergence. In Appendix \ref{AppendixB} we demonstrate how the algorithm’s convergence depends on this exponent for graph instances relevant to practical electronic structure simulations.


\subsection{Graph Updating}
In order to propose a candidate $\mathcal{G}'$ deriving from $\mathcal{G}$, we require an updating function. In particular, $\mathcal{G}'$ should result from a given relabelling function $\mathcal{M}^k_t$ of depth $k$. To encode such relabelling function, we create a list of length $k$, where the $i$'th entry of the list will encode the swapping information of the $i$'th layer of the swapping relabelling function. Such information, is a subset of the edges of $\mathcal{G}$, where no label is repeated. 
\\
\\
The graph updating chooses a random layer from 1 to $k$. Then with probability $p$ adds a random edge that is not appearing in that particular layer, or with probability $1-p$ removes a random edge from that layer.

\subsection{The Algorithm}
The procedure starts with the initial graph $\mathcal{G}^{t=0}$ and the corresponding history matrix $H^{t=0}$. The algorithm relabels the graph with a swap depth $k$, yielding candidate graphs $\mathcal{G}^{t+1}_{candidate} = \mathcal{M}^k_t(\mathcal{G}^{t})$, with associated candidate history matrices $H_{candidate}^{t+1}$. The candidates are optimized through a simulating annealing protocol in order to minimize the cost function $\mathcal{C}$ in Equation \ref{CostFunction}. Multiple annealing schedules may be run in parallel to increase the optimization quality. The best candidate found along the annealing is promoted as $\mathcal{G}^{t+1}$, as well as its associated history matrix $H^{t+1}$. The algorithm terminates when every label has been a neighbour of every other label, i.e., when $H_{i,j} = 0$ for all $i,j$. The algorithm is detailed in Alg.\ref{GraphAlgorithm}. A visual example of a first optimization step for $k=2$ is shown in Fig.\ref{fig:SimulatedAnnealing}.
\\
\\
Since the algorithm iterates until the history matrix is fully cleared, the required number of iterations scales with the graph’s average degree, i.e., the mean number of edges per node:
\begin{equation}
    \hat{D}(G)= \frac{2|E|}{|V|}.
\end{equation}
 Each qubit creates \(\mathcal{O}(\hat{D}(G))\) new adjacencies per iteration on average. As each must connect to the other $N-1$ qubits, to clear the history matrix the total number of iterations scales as
 \begin{equation}
    \mathcal{O}\!\left(\frac{N}{\hat{D}(G)}\right).
\end{equation}
In practical hardware of interest, the average degree \(\hat{D}(G)\) is typically \(\mathcal{O}(1)\). Thus, the algorithm does not change the asymptotic scaling of SWAP-routing depth with system size, which remains linear in \(N\). Its purpose is instead to exploit the fact that experimental devices often have connectivity larger than a simple line, \(\hat{D}(G)>2\), to reduce the constant prefactor: higher degree creates more new adjacencies per iteration on average, thereby reducing the overall SWAP-network depth compared to the strictly linear case.
\\
\\
The simulated-annealing hyperparameters control the trade-off between runtime and solution quality. Since the routing/ordering task is a hard combinatorial problem, we typically seek good (not necessarily optimal) schedules. Larger computational budgets usually improve the result, while smaller budgets still provide adequate schedules with efficient scaling.
\\
\\
For the small quantum circuits considered in this work, the algorithm is always able to terminate after a modest computational effort and swap network depth, yielding the complete swap network where all the qubits connect. It is however expected that for much larger circuits it is harder to attain a situation where $H_{i,j} = 0$ for all $i,j$, and that the required swap network depth exceeds the noise-limited circuit depth that can be implemented in hardware. In those cases, a stopping criterion may be applied after a given number of iterations or connections in the history matrix ($I_{max}$), yielding a partial swap network.


\begin{algorithm}[H]
  \caption{Swap network optimization under arbitrary constrained qubit connectivity}
  \begin{algorithmic}[1]
  \Require $\mathcal{G} = (V, E, \mathcal{L})$\Comment{Labelled connectivity graph} 
  \Require  $k$ \Comment{Swap depth}
  \Require  $I_{max}$ \Comment{Max iterations (optional)}
    \State \( H_{i,j} \equiv
        \begin{cases}
        0 & \text{if } (v_i, v_j) \in E \text{ or } i = j, \\
        1 & \text{otherwise}.
        \end{cases} \) 
    \State SwapProtocol=[$\hspace{2pt}$]
    \While{$\exists (i,j) \text{such that } H_{i,j}\neq 0 \;\land\;\lvert SwapProtocol\rvert < I_{\max}$}

    \State Optimize $\mathcal{R}^k(\mathcal{G})$, $H'$, minimizing $\mathcal{C}(\mathcal{R}^k(\mathcal{G}),H')$
    \State $\mathcal{G} \xleftarrow{} \mathcal{R}^k (\mathcal{G})$ 
    \State $H \xleftarrow{} H'$
    \State Append $\mathcal{R}^k$ to SwapProtocol
    \EndWhile
    
    \Return SwapProtocol
  \end{algorithmic}
  \label{GraphAlgorithm}
\end{algorithm}

\begin{figure*} [t]                    
\centering
\includegraphics[width=2\columnwidth, angle=0]{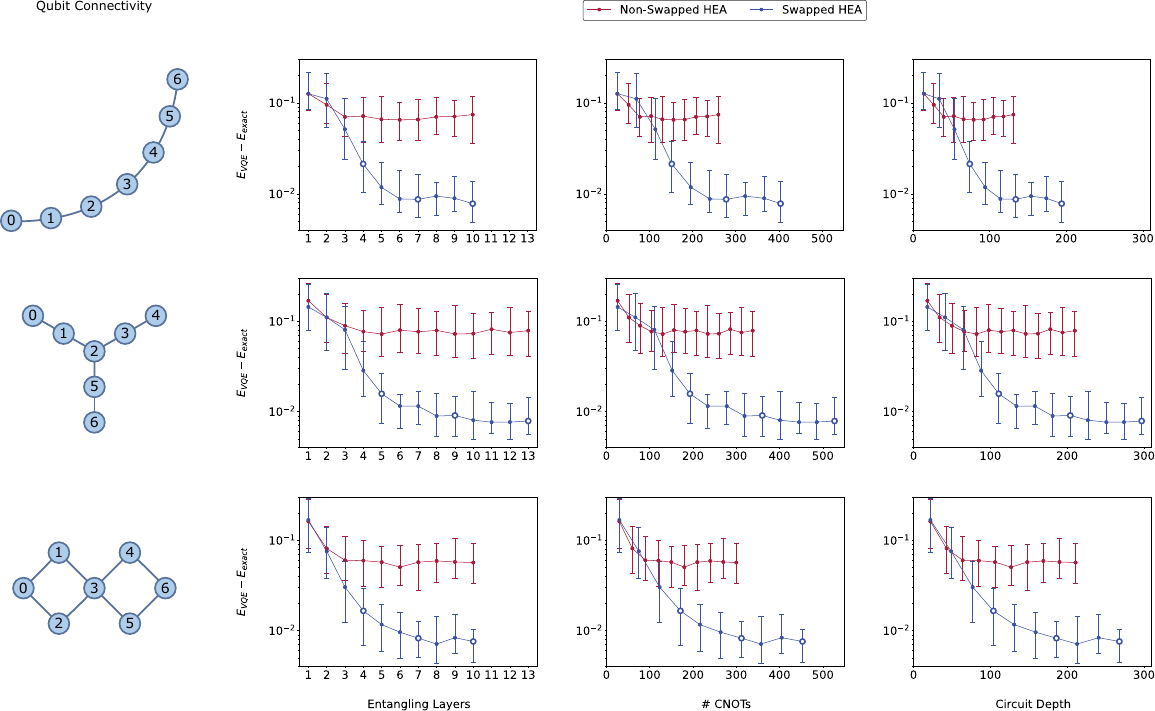}
\caption{VQE energy error statistics for 100 random 7-spin glass instances, evaluated across varying qubit connectivities and relevant parameters. We compare the standard HEA to the version with optimized swap layers added between entangling layers across three connectivity configurations: linear (top row), heavy-hex (middle row), and square-lattice (bottom row). The energy error is shown as a function of entangling layers, CNOT count and circuit depth. CNOT count and circuit depth include the decomposition of swap gates as 3 CNOTs, and CRy with 2 CNOTs. Median values across the 100 instances are represented by dots, with error bars indicating interquartile ranges. Highlighted blue circles denote the complete application of the optimized swap network once, twice, and three times sequentially.
}
\label{fig:Spinglass}
\end{figure*}

\subsection{Ansatz Design}
Once an optimized swap network is found for a given qubit connectivity using the proposed algorithm, embedding a layered ansatz within it becomes straightforward. The entangling layers are interleaved with the swap layers from the optimized swap network, as displayed in Fig.\ref{fig:Infographic}. 

\section{Numerical Results}
To evaluate how optimized swap networks enhance optimization in systems with long-range correlations that do not align with hardware connectivity, in the following we conduct a VQE simulation on highly correlated systems. All simulations were implemented in \textsc{tequila} \cite{kottmann2021tequila}, using \textsc{qulacs} simulator \cite{suzuki2021qulacs}.

\subsection{Spin Glass Simulations}
In this section, we simulate the quantum ground state of random spin-glass models with all-to-all interactions. These models generate highly correlated ground states, where every subparty shares strong correlations with all others. The Hamiltonian is given by:
\[
\hat{H} = \sum_{i<j} J_{ij} \hat{\sigma}_i^x \hat{\sigma}_j^x + \sum h_i \hat{\sigma}^z_i
\]
With, \( J_{ij} \) interaction coefficients, \( h_i \) on-site magnetic fields, and \( \hat{\sigma}_i^x \), \( \hat{\sigma}_i^z \) the Pauli-X and Pauli-Z operators on spin \( i \). Uniformly distributed random values between -1 and 1 are assigned for both \( J_{ij} \) and \( h_i \).
\\
\\
To perform simulation statistics, we generate 100 different random spin-glass instances, each with $N=7$ spins. For each case, we perform a VQE optimization on a vanilla setting for the HEA, which we refer to as CRy-HEA. In this setup, each two-qubit entangling gate consists of an Ry gate applied to each qubit, followed by a CRy gate, with each gate having its own parameter. Details on the implementation of CRy-HEA for different connectivities are found on Appendix \ref{AppendixB}. The resulting energy statistics are compared to the exact ground-state energy found with exact diagonalization, across different depths of the CRy-HEA, both with and without the swap protocol, and for various qubit connectivity configurations, see Fig.~\ref{fig:Spinglass}. For both the swapped and non-swapped versions of the HEA, CRy and swap gates are decomposed into native gates, consisting of single-qubit gates and CNOTs. Additionally, the VQE energy precision is shown as a function of the total number of CNOTs (major source of noise) and the overall circuit depth (qubit decoherence). 
\\
\\
Ground-state optimization of random spin glasses is challenging due to the complex energy landscape, which features many local minima that can trap optimizers. Here, we used the gradient-free COBYLA optimizer, which mitigates the effect of getting trapped in shallow local minima. The optimization was set up to 10000 iterations.
\\
\\
Although highly expressive at deep circuits, we observe that the Ry-HEA optimization without swaps tends to saturate at a certain energy. This occurs because the Ry-HEA cannot capture the necessary correlations between long-range qubits, leading to an insufficient representation of the ground state. However, when the Ry-HEA is embedded into an optimized swap network tailored to the given connectivity, it can more effectively capture correlations between all qubits in the system. This leads to a better representation of the ground state and, consequently, a reduction in the error of the ground state energy.
\\
\\
When comparing energy errors as a function of the parameters, the HEA embedded within the swap network yields better results for a fixed number of CNOTs, fixed circuit depth, or number of optimization parameters. Therefore, although the additional swap layers introduce an overhead in terms of CNOTs, circuit depth and parameters, their use proves beneficial to the optimization accuracy. Note that this translates into higher noise efficiency: for a fixed circuit depth or CNOT count (therefore comparable accumulated noise), it achieves lower errors and more reliable convergence.



\subsection{Electronic Structure Simulations}
\begin{figure}[b]
    \centering
    \includegraphics[width=1\linewidth]{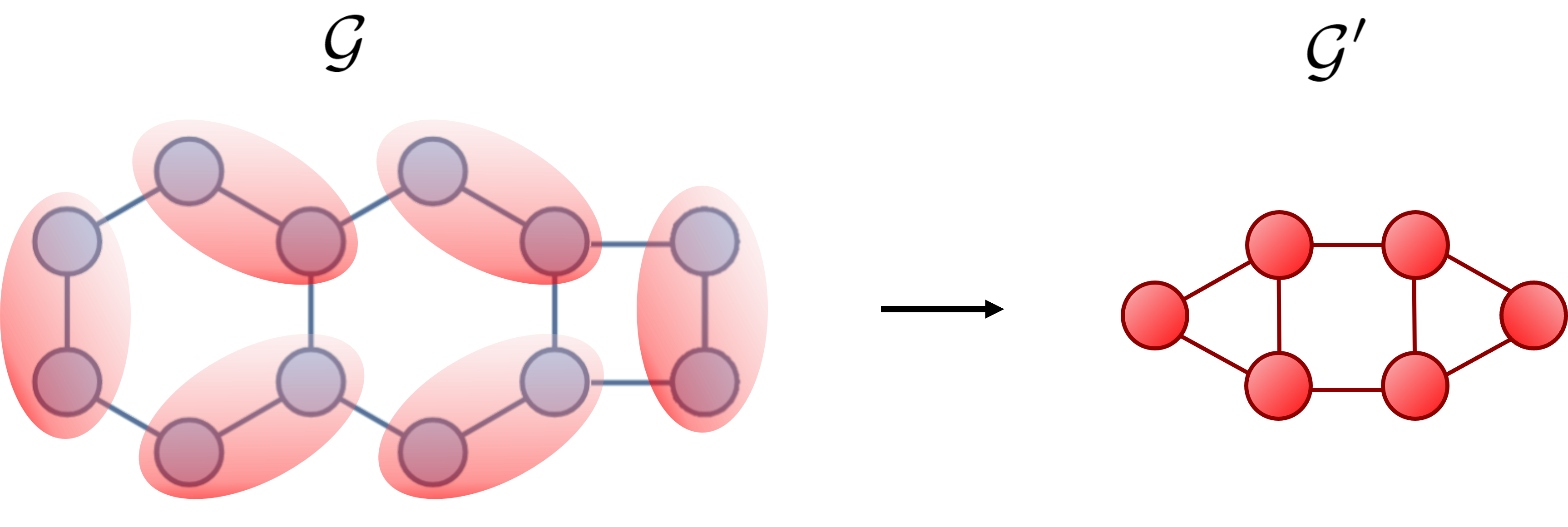}
    \caption{Example of connectivity graph coarsening from the spin-orbital (qubit) level $\mathcal{G}$  to the molecular orbital level $\mathcal{G}'$. Each blue node corresponds to a qubit holding the occupancy information of a single spin orbital. Each red node corresponds to a pair of qubits holding the occupancy information of a molecular orbital.}
    \label{fig:COnnectivityCoarsening}
\end{figure}
\begin{figure*} [t]                    
\centering
\includegraphics[width=2\columnwidth, angle=0]{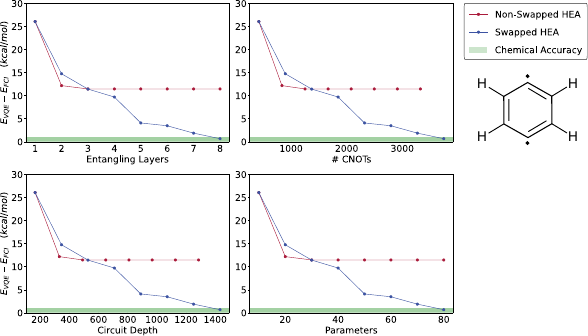}
\caption{VQE energy error for the $\pi$ orbital system of p-benzyne birradical (Appendix.\ref{AppendixB}) on the qubit connectivity of [Fig. \ref{fig:COnnectivityCoarsening}]. The ansatz optimization with all initialization angles close to zero (in red), is compared to the case where the swap network is embedded (in blue). The energy error is shown as a function of entangling layers, CNOT count, circuit depth, and optimization parameters.
}
\label{fig:pBenzyne_Opt}
\end{figure*} 
We now turn to problems of real-world relevance, consisting in the electronic structure problems. These are central to quantum chemistry and represent the foundation of transversal theoretical simulations in fields such as drug discovery \cite{doi:10.1021/acs.jctc.2c00574, 8585034}, computational catalysis \cite{von_Burg_2021}, or materials science \cite{Lordi_Nichol_2021}. 
VQAs provide a promising way to address those physically structured problems, both in near-term and long-term quantum hardware \cite{zimborás2025mythsquantumcomputationfault}. The proposed method is readily extendable to these physically structured systems, as demonstrated in this section.
\\
\\
The eigenstates of the electronic‑structure Hamiltonian are the key objects underpinning most analysis in quantum chemistry. In brief, a Hamiltonian expressed in the Fermionic basis can be translated into an equivalent qubit operator by applying a suitable Fermion‑to‑qubit encoding. Among many of these available encodings \cite{Bravyi2002,PRXQuantum.5.030333, algaba2025fermiontoqubitencodingsarbitrarycode, Vlasov_2022, miller2022bonsai,miller2024treespilationarchitecturestateoptimisedfermiontoqubit,chiew2025optimalfermionqubitmappingsquadratic,chiew2024ternarytreetransformationsequivalent,harrison2024sierpinskitrianglefermiontoqubittransform}, the Jordan-Wigner transformation \cite{jordan1993paulische} represents the paradigmatic one, encoding the occupation of spin-orbitals locally in the qubits: the $i$th qubit is in state $|0\rangle$ if the $i$th spin-orbital is unoccupied, and $|1\rangle$ if it is occupied.
\\
\\
A typical reference state for variational optimization consists in the Hartree-Fock state, which is the single determinant with minimum energy. The Unitary Coupled Cluster (UCC) framework parametrizes the Hartree Fock reference state via single and double excitations, resticting the parametrization to be particle-preserving \cite{Coester_Kümmel_1960b}, see Appendix \ref{AppendixA}. An effective ansatz variant consists in the Unitary paired Coupled Cluster of generalized Singles and Doubles (UpCCGSD), which restricts excitations to occur between spin-orbital pairs belonging to the same molecular orbital \cite{Lee_Huggins_Head-Gordon_Whaley_2018, Anand_Schleich_Alperin-Lea_Jensen_Sim_Díaz-Tinoco_Kottmann_Degroote_Izmaylov_Aspuru-Guzik_2022}. This significantly reduces the resource complexity by limiting operations to molecular orbital interactions instead of arbitrary spin-orbital excitations. Here, we use approximations of these single and double paired excitations as the entangling operations (integrated in \textsc{tequila} \cite{kottmann2021tequila}), and embed them within a Fermionic swap network that routes the molecular orbital basis through the processor's connectivity. Details on the implementation of the approximate single and double excitations in arbitrary qubit connectivity, can be found in Appendix \ref{AppendixA} .
\\
\\
The qubits representing the two spin-orbitals from the same molecular orbital are positioned adjacently in the initial qubit connectivity graph $\mathcal{G}$ (see [Fig.\ref{fig:COnnectivityCoarsening}]). Such qubit pairs can be coarsened to yield a connectivity graph $\mathcal{G}'$ in the molecular orbital picture, indicating which pairs of molecular orbitals can be directly correlated in the processor. It is important to note that the graphs $\mathcal{G}$ and $\mathcal{G}'$ discussed here are unrelated to valence bond graphs used in quantum chemistry and other ansatz design approaches \cite{kottmann2023molecular}. While valence bond graphs represent electron pairings and bonding topologies in a molecule, the connectivity graphs in this context describe the physical or logical arrangement of qubits and how molecular orbitals are mapped onto them for quantum simulation purposes. Within this picture, the approximate implementation of single and double excitation operations serve as building blocks for entangling layers (analogous to the HEA case), which correlate adjacent molecular orbitals in $\mathcal{G}'$. Such entangling layers are embedded within an optimized Fermionic swap network on the molecular orbital connectivity $\mathcal{G}'$ using Alg.\ref{GraphAlgorithm}. Implementation details can be found in Appendix \ref{AppendixB} and in the source code \cite{AutoSwap}.
\\
\\
We simulate the ground state of the $\pi$ system of p-benzyne (see [Fig.\ref{fig:Benzene_Orbitals}]) using the cc-pVDZ basis set (12 spin-orbitals/qubits), detailed in Appendix \ref{AppendixB}. The system exhibits an open-shell singlet biradical ground state, corresponding to a challenging strongly correlated state. We use the L-BFGS-B optimizer \cite{doi:10.1137/0916069} up to a maximum of 1000 iterations. In the example connectivity from Fig.\ref{fig:COnnectivityCoarsening}, we compare the performance of the non-swapped ansatz version, with the swapped one including the Fermionic swap layers in between. As for a given amount of entangling layers the swapped version has additional resources due to the additional Fermionic swaps. We decompose the Fermionic swap operations and count the resulting elementary CNOT gates, circuit depth, and variational parameters for each simulated circuit, enabling a fair comparison between the ansätze. The converged VQE energies are also displayed as a function of the relevant parameters. All the optimizations have been performed with optimization angles close to zero \cite{hibatallah2023frameworkdemonstratingpracticalquantum,Grant_2019}, motivated by the swap network initialization. Alternatively, the swap network can be incorporated in certain cases into the initial parameters of the optimization, as single and double excitations with angles of $\pi$ effectively act as Fermionic swaps \cite{Anselmetti_Wierichs_Gogolin_Parrish_2021}. We leave the exploration of this integration for arbitrary connectivity to future work.
\\
\\
Numerical results are displayed in Fig.\ref{fig:pBenzyne_Opt}. We observe that the non-swapped version of the ansatz struggles to achieve chemical precision. Moreover, increasing the circuit depth up to the eight entangling layers tested, does not lead to improved accuracy. In contrast, as seen in the previous section, the swapped version of the ansatz guarantees efficient trainability, converging rapidly to the solution as simulation resources are increased. When analyzing the resulting circuits in terms of CNOT count, circuit depth, and number of optimization parameters, we find that the swapped ansatz consistently outperforms its counterpart in all of the variables. As for spin glasses, this again shows higher resilience to noise. With the same depth or number of entangling gates, and therefore equivalent noise levels, it provides improved convergence. Ultimately, this highlights the advantages of incorporating swap networks into the ansatz for practical noisy simulations.
\subsection{Scaling Analysis}
\begin{figure}[h!]
    \centering
    \includegraphics[width=1\linewidth]{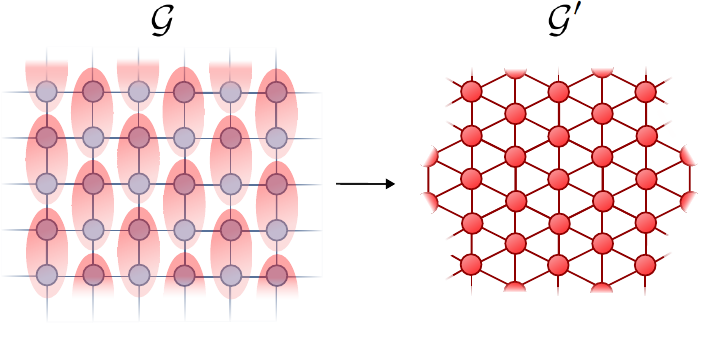}
    \caption{Selected square-grid connectivity graph coarsening from the spin-orbital (qubit) level $\mathcal{G}$ to the molecular-orbital level $\mathcal{G}'$. The resulting effective graph $\mathcal{G}'$ exhibits hexagonal symmetry with additional nodes at the centers of the hexagonal plaquettes.}
    \label{fig:SquareGridCoarsening}
\end{figure}
\begin{figure}[h!]
    \centering
    \includegraphics[width=1\linewidth]{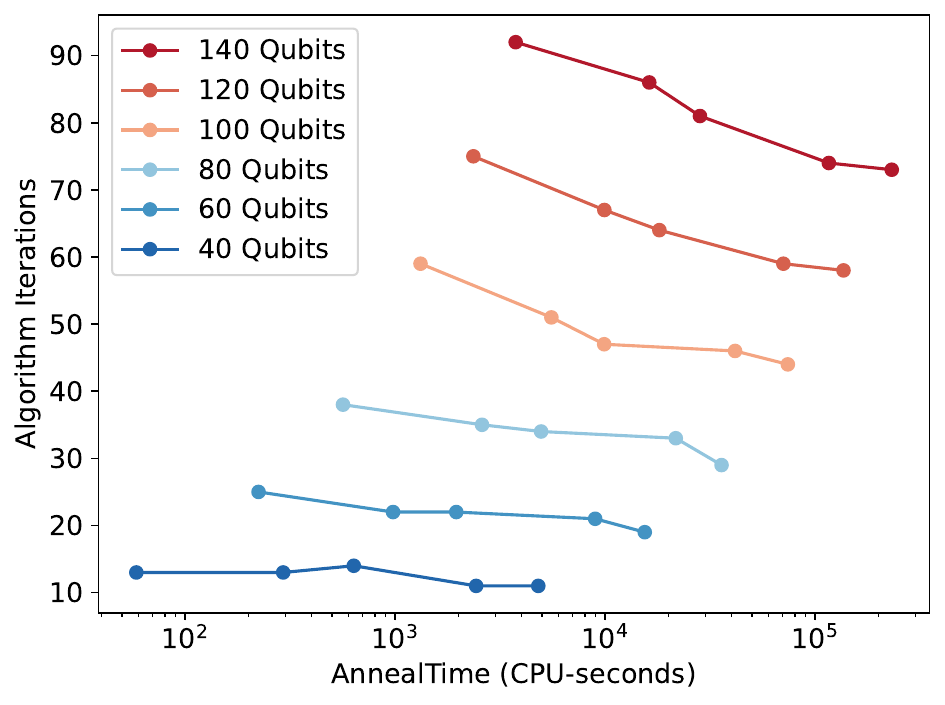}
    \caption{Scaling of algorithm iterations with qubit count and annealing time. Graph instances are from the family in Fig.\ref{fig:SquareGridCoarsening}, generated with the function $\mathit{create\_hex\_triangular\_grid}$. }
    \label{fig:Nq_vs_Iterations}
\end{figure}
We now evaluate the scaling of the algorithm in the context of practical electronic-structure simulations on realistic quantum processors. Modern quantum processors typically feature two-dimensional square-grid connectivity, which we will now adapt to electronic-structure problems. Following the procedure introduced in the previous section, we apply the graph coarsening illustrated in Fig.\ref{fig:SquareGridCoarsening}, which maps the native qubit-level connectivity $\mathcal{G}$ onto an effective orbital-level interaction graph $\mathcal{G}'$. The resulting graph $\mathcal{G}'$ exhibits hexagonal symmetry with additional nodes at the centers of the hexagonal plaquettes. This connectivity is particularly advantageous, as it features enhanced coordination numbers: bulk nodes are connected to six neighbors, while boundary nodes exhibit between two and five connections. In the source code \cite{AutoSwap}, we provide the function $\mathit{create\_hex\_triangular\_grid}$ that constructs graphs from this family for a prescribed number of qubits (see examples in Appendix \ref{AppendixB}).
\\
\\
 We evaluate the scaling of the classical optimization algorithm on these instances as a function of the graph size in Fig.\ref{fig:Nq_vs_Iterations}. 
 The number of algorithm iterations increases with larger instances. This is expected, as more complex swap protocols and deeper swap networks are required.
\\
\\
As already discussed, the scaling remains practically efficient because annealing time sets a clear trade off between runtime and solution quality. Short anneals produce schedules quickly and scale to large instances, but they typically give lower quality solutions and deeper swap networks. Longer anneals cost more, yet they more reliably find higher quality schedules, often yielding shallower and more efficient swap networks. Therefore, even for large system sizes, one can choose fast approximate solutions or slower higher quality solutions depending on the available runtime budget.
\subsection{Comparison to Dynamical Routing}
Finally, we compare the swap network augmented ansatz with a dynamically routed all-to-all ansatz generated by a round-robin schedule and compiled with SABRE~\cite{li2019tacklingqubitmappingproblem, zou2024lightsabrelightweightenhancedsabre}.
\begin{figure}[h!]
    \centering
    \includegraphics[width=1\linewidth]{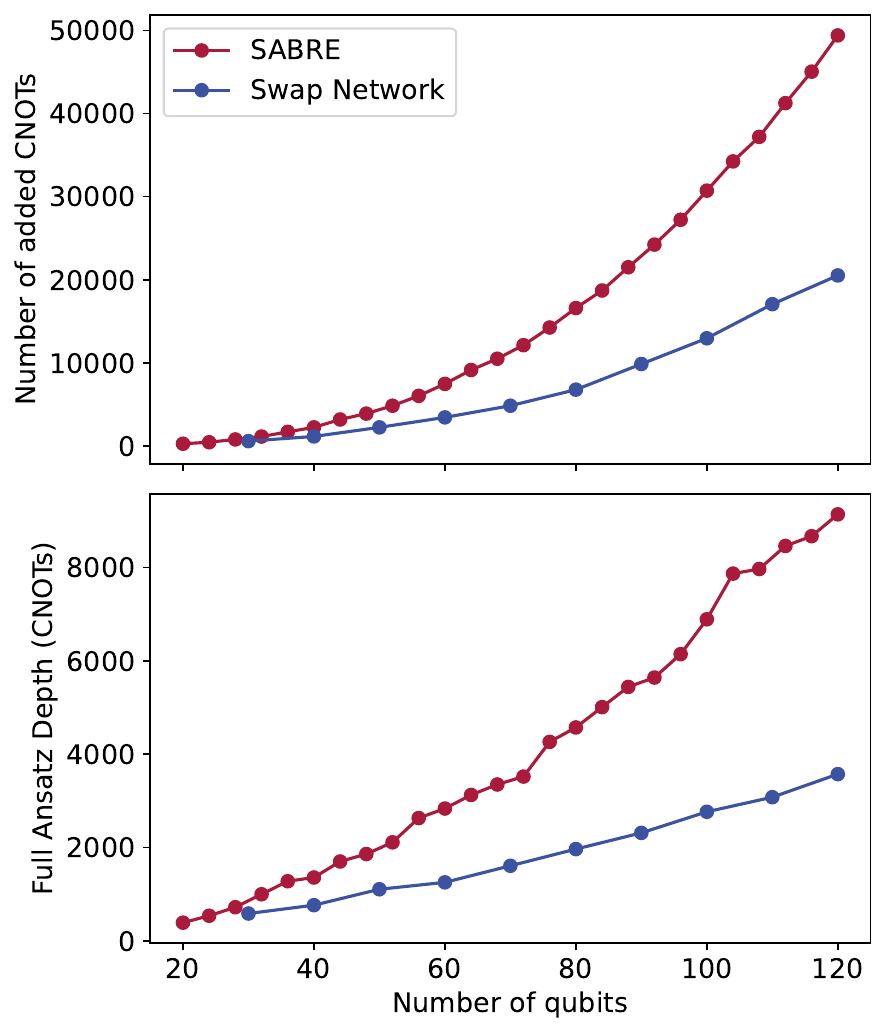}
    \caption{Comparison of circuit depth and added FSWAP count to implement a full connectivity ansatz for SABRE and the proposed swap network augmented ansatz. The comparison is conducted across instances of different graph sizes generated by the function $\mathit{create\_hex\_triangular\_grid}$.}
    \label{fig:SabreComparison}
\end{figure}
For both approaches, and for different graph sizes within the family shown in Fig.~\ref{fig:SquareGridCoarsening}, we decompose the circuits into native gates and report the resulting total CNOT depth in Fig.~\ref{fig:SabreComparison}. We also report the total number of fSWAP gates inserted by each method, namely SABRE routing and the swap network augmented construction.
\\
\\
The swap network augmented ansatz requires fewer fSWAP gates in the final circuit, and since each fSWAP decomposes into multiple CNOT gates, this directly reduces the entangling gate overhead and yields a significantly shallower circuit in CNOT depth. This improvement stems from how the ansatz is constructed. It is designed to exploit the native device connectivity from the start, with a swap network that optimizes the swap protocol while the entangling pattern is generated directly from the available couplings, so much of the routing is handled by the ansatz design itself. In contrast, an all to all ansatz is defined at the logical level and must later be embedded into the hardware graph, typically requiring longer and more complex swap protocols. SABRE is highly effective at compiling such abstract circuits to a given connectivity, but the swap network approach reduces the need for heavy routing by avoiding an all-to-all target and instead constructing the required connectivity directly on the hardware.
\section{Outlook}
Efficient design of parametrized circuit ansätze is key to the performance of variational quantum optimization. Suitable initialization techniques and parametrizations that allow an effective navigation to the solution state are a major concern. While ansätze adapted to the connectivity (eg. HEA, naive UpCCGSD) can significantly reduce resources, here we have observed that their trainability is severely compromised when using an unstructured and agnostic flow of information within the ansatz, getting stuck in a local minima.
\\
\\
To overcome these problems, we have proposed to include swap networks in the optimisation, which allow overcoming such trainability barriers in non-local correlated systems, as arbitrary pairs of qubits can be directly correlated in the ansatz. Although additional resources are required for building the swap network, for the same amount of total resources the overall performance of the optimization algorithm is improved, showcasing the advantage of a synergystic hardware-friendly ansatz with a swap network. Key in our construction is the algorithm that allows the classical optimization for the swap network given an arbitrary qubit connectivity, making the swap-augmented ansatz readily available for variational optimization in the devices.  Importantly, our algorithm is architecture-agnostic: it adapts to any qubit layout, and can skip faulty qubits by excluding them from the connectivity graph. Ultimately, we set optimized swap networks as a tool to be exploited in ansatz design, bridging the gap between hardware constraints and algorithmic performance.
\\
\\
Future work may explore swap network augmented ansätze in other contexts, such as in quantum machine learning, where circuit trainability and efficient parameterization are equally critical.Beyond simulation, it would be valuable to validate these results on real quantum hardware. More broadly, our approach has implications for beneficial circuit compilation under constrained connectivity, suggesting routing/compilation strategies that trade additional structure for improved effective expressivity. Future work may explore these directions in practical compilation and experimental demonstrations. Finally, our results open new avenues in the understanding of the tradeoff between trainability and expressivity of variational quantum circuits that deserve further investigation. In this sense, note that, when comparing circuits of equal depth, there is a choice of gate parameters in the circuit without SWAP networks that may allow recovering the circuit with the SWAP networks. That is, the ansatz without these networks is in principle more expressive than the one including them. Yet, the trainability is much harder and, in particular, a very subtle fine tuning of the parameters is required to obtain the ansatz with the SWAP networks. This implies that the parameter exploration in the optimisation is not able to find the solution with the SWAP networks and the circuit performance is in practice much worse, despite being more expressive in theory. 
\\
\section*{ACKNOWLEDGEMENTS}
 We thank Júlia Barberà-Rodríguez for discussions. This work is supported by the ERC AdG CERQUTE, the Government of Spain (FUNQIP, NextGenerationEU PRTRC17.I1 and Quantum in Spain, Severo Ochoa CEX2019-000910-S), Fundació Cellex, Fundació Mir-Puig, Generalitat de Catalunya (CERCA programme), the AXA Chair in Quantum Information Science, EU project PASQUANS2, QSNP 101114043, NeQST 101080086 and COMPUTE PCI2024-153430. Ayuda PREP2022-000452 financiada por MCIN/AEI/10.13039/501100011033 y por el FSE+.
\bibliographystyle{quantum}
\bibliography{lit}

\begin{thebibliography}{10}

\bibitem{Cerezo_Arrasmith_Babbush_Benjamin_Endo_Fujii_McClean_Mitarai_Yuan_Cincio}
M.~Cerezo, Andrew Arrasmith, Ryan Babbush, Simon~C. Benjamin, Suguru Endo, Keisuke Fujii, Jarrod~R. McClean, Kosuke Mitarai, Xiao Yuan, Lukasz Cincio, and et~al.
\newblock ``Variational quantum algorithms''.
\newblock \href{https://dx.doi.org/10.1038/s42254-021-00348-9}{Nature Reviews Physics {\bf 3}, 625–644}~(2021).

\bibitem{McClean_Romero_Babbush_Aspuru-Guzik_2016}
Jarrod~R McClean, Jonathan Romero, Ryan Babbush, and Alán Aspuru-Guzik.
\newblock ``The theory of variational hybrid quantum-classical algorithms''.
\newblock \href{https://dx.doi.org/10.1088/1367-2630/18/2/023023}{New Journal of Physics {\bf 18}, 023023}~(2016).

\bibitem{Tilly_Chen_Cao_Picozzi_Setia_Li_Grant_Wossnig_Rungger_Booth}
Jules Tilly, Hongxiang Chen, Shuxiang Cao, Dario Picozzi, Kanav Setia, Ying Li, Edward Grant, Leonard Wossnig, Ivan Rungger, George~H. Booth, and et~al.
\newblock ``The variational quantum eigensolver: A review of methods and best practices''.
\newblock \href{https://dx.doi.org/10.1016/j.physrep.2022.08.003}{Physics Reports {\bf 986}, 1–128}~(2022).

\bibitem{Sim_Johnson_Aspuru‐Guzik_2019}
Sukin Sim, Peter~D. Johnson, and Alán Aspuru‐Guzik.
\newblock ``Expressibility and entangling capability of parameterized quantum circuits for hybrid quantum‐classical algorithms''.
\newblock \href{https://dx.doi.org/10.1002/qute.201900070}{Advanced Quantum Technologies{\bf 2}}~(2019).

\bibitem{romero2018strategiesquantumcomputingmolecular}
Jonathan Romero, Ryan Babbush, Jarrod~R. McClean, Cornelius Hempel, Peter Love, and Alán Aspuru-Guzik.
\newblock ``Strategies for quantum computing molecular energies using the unitary coupled cluster ansatz''~(2018).
\newblock  \href{http://arxiv.org/abs/1701.02691}{arXiv:1701.02691}.

\bibitem{Kandala_2017}
Abhinav Kandala, Antonio Mezzacapo, Kristan Temme, Maika Takita, Markus Brink, Jerry~M. Chow, and Jay~M. Gambetta.
\newblock ``Hardware-efficient variational quantum eigensolver for small molecules and quantum magnets''.
\newblock \href{https://dx.doi.org/10.1038/nature23879}{Nature {\bf 549}, 242--246}~(2017).

\bibitem{li2019tacklingqubitmappingproblem}
Gushu Li, Yufei Ding, and Yuan Xie.
\newblock ``Tackling the qubit mapping problem for nisq-era quantum devices''~(2019).
\newblock  \href{http://arxiv.org/abs/1809.02573}{arXiv:1809.02573}.

\bibitem{https://doi.org/10.4230/lipics.tqc.2019.5}
Alexander Cowtan, Silas Dilkes, Ross Duncan, Alexandre Krajenbrink, Will Simmons, and Seyon Sivarajah.
\newblock ``On the qubit routing problem''~(2019).

\bibitem{murali2019noiseadaptivecompilermappingsnoisy}
Prakash Murali, Jonathan~M. Baker, Ali~Javadi Abhari, Frederic~T. Chong, and Margaret Martonosi.
\newblock ``Noise-adaptive compiler mappings for noisy intermediate-scale quantum computers''~(2019).
\newblock  \href{http://arxiv.org/abs/1901.11054}{arXiv:1901.11054}.

\bibitem{zhu2025quantumcompilerdesignqubit}
Chenghong Zhu, Xian Wu, Zhaohui Yang, Jingbo Wang, Anbang Wu, Shenggen Zheng, and Xin Wang.
\newblock ``Quantum compiler design for qubit mapping and routing: A cross-architectural survey of superconducting, trapped-ion, and neutral atom systems''~(2025).
\newblock  \href{http://arxiv.org/abs/2505.16891}{arXiv:2505.16891}.

\bibitem{Holmes_Johri_Guerreschi_Clarke_Matsuura_2020}
Adam Holmes, Sonika Johri, Gian~Giacomo Guerreschi, James~S Clarke, and A~Y Matsuura.
\newblock ``Impact of qubit connectivity on quantum algorithm performance''.
\newblock \href{https://dx.doi.org/10.1088/2058-9565/ab73e0}{Quantum Science and Technology {\bf 5}, 025009}~(2020).

\bibitem{McClean_2018}
Jarrod~R. McClean, Sergio Boixo, Vadim~N. Smelyanskiy, Ryan Babbush, and Hartmut Neven.
\newblock ``Barren plateaus in quantum neural network training landscapes''.
\newblock \href{https://dx.doi.org/10.1038/s41467-018-07090-4}{Nature Communications{\bf 9}}~(2018).

\bibitem{PhysRevLett.120.110501}
Ian~D. Kivlichan, Jarrod McClean, Nathan Wiebe, Craig Gidney, Al\'an Aspuru-Guzik, Garnet Kin-Lic Chan, and Ryan Babbush.
\newblock ``Quantum simulation of electronic structure with linear depth and connectivity''.
\newblock \href{https://dx.doi.org/10.1103/PhysRevLett.120.110501}{Phys. Rev. Lett. {\bf 120}, 110501}~(2018).

\bibitem{ogorman2019generalizedswapnetworksnearterm}
Bryan O'Gorman, William~J. Huggins, Eleanor~G. Rieffel, and K.~Birgitta Whaley.
\newblock ``Generalized swap networks for near-term quantum computing''~(2019).
\newblock  \href{http://arxiv.org/abs/1905.05118}{arXiv:1905.05118}.

\bibitem{Anselmetti_Wierichs_Gogolin_Parrish_2021}
Gian-Luca~R Anselmetti, David Wierichs, Christian Gogolin, and Robert~M Parrish.
\newblock ``Local, expressive, quantum-number-preserving vqe ansätze for fermionic systems''.
\newblock \href{https://dx.doi.org/10.1088/1367-2630/ac2cb3}{New Journal of Physics {\bf 23}, 113010}~(2021).

\bibitem{hagge2022optimalfermionicswapnetworks}
Tobias Hagge.
\newblock ``Optimal fermionic swap networks for hubbard models''~(2022).
\newblock  \href{http://arxiv.org/abs/2001.08324}{arXiv:2001.08324}.

\bibitem{PhysRevResearch.4.033028}
Akel Hashim, Rich Rines, Victory Omole, Ravi~K. Naik, John~Mark Kreikebaum, David~I. Santiago, Frederic~T. Chong, Irfan Siddiqi, and Pranav Gokhale.
\newblock ``Optimized swap networks with equivalent circuit averaging for qaoa''.
\newblock \href{https://dx.doi.org/10.1103/PhysRevResearch.4.033028}{Phys. Rev. Res. {\bf 4}, 033028}~(2022).

\bibitem{PhysRevResearch.6.043279}
J\'ulia Barber\`a-Rodr\'{\i}guez, Nicolas Gama, Anand~Kumar Narayanan, and David Joseph.
\newblock ``Finding dense sublattices as low energy states of a hamiltonian''.
\newblock \href{https://dx.doi.org/10.1103/PhysRevResearch.6.043279}{Phys. Rev. Res. {\bf 6}, 043279}~(2024).

\bibitem{AutoSwap}
Teodor Parella-Dilmé.
\newblock ``Autoswap''.
\newblock \url{https://github.com/teoparella/AutoSwap}~(2026).

\bibitem{zou2024lightsabrelightweightenhancedsabre}
Henry Zou, Matthew Treinish, Kevin Hartman, Alexander Ivrii, and Jake Lishman.
\newblock ``Lightsabre: A lightweight and enhanced sabre algorithm''~(2024).
\newblock  \href{http://arxiv.org/abs/2409.08368}{arXiv:2409.08368}.

\bibitem{Grimsley_Economou_Barnes_Mayhall_2019}
Harper~R. Grimsley, Sophia~E. Economou, Edwin Barnes, and Nicholas~J. Mayhall.
\newblock ``An adaptive variational algorithm for exact molecular simulations on a quantum computer''~(2019).

\bibitem{PRXQuantum.2.020310}
Ho~Lun Tang, V.O. Shkolnikov, George~S. Barron, Harper~R. Grimsley, Nicholas~J. Mayhall, Edwin Barnes, and Sophia~E. Economou.
\newblock ``Qubit-adapt-vqe: An adaptive algorithm for constructing hardware-efficient ans\"atze on a quantum processor''.
\newblock \href{https://dx.doi.org/10.1103/PRXQuantum.2.020310}{PRX Quantum {\bf 2}, 020310}~(2021).

\bibitem{Yordanov_Armaos_Barnes_Arvidsson-Shukur_2021}
Yordan~S. Yordanov, V.~Armaos, Crispin~H. Barnes, and David~R. Arvidsson-Shukur.
\newblock ``Qubit-excitation-based adaptive variational quantum eigensolver''.
\newblock \href{https://dx.doi.org/10.1038/s42005-021-00730-0}{Communications Physics{\bf 4}}~(2021).

\bibitem{Larocca_Thanasilp_Wang_Sharma_Biamonte_Coles_Cincio_McClean_Holmes_Cerezo_2025}
Martín Larocca, Supanut Thanasilp, Samson Wang, Kunal Sharma, Jacob Biamonte, Patrick~J. Coles, Lukasz Cincio, Jarrod~R. McClean, Zoë Holmes, and M.~Cerezo.
\newblock ``Barren plateaus in variational quantum computing''.
\newblock \href{https://dx.doi.org/10.1038/s42254-025-00813-9}{Nature Reviews Physics {\bf 7}, 174–189}~(2025).

\bibitem{Leone2024practicalusefulness}
Lorenzo Leone, Salvatore~F.E. Oliviero, Lukasz Cincio, and M.~Cerezo.
\newblock ``On the practical usefulness of the {H}ardware {E}fficient {A}nsatz''.
\newblock \href{https://dx.doi.org/10.22331/q-2024-07-03-1395}{{Quantum} {\bf 8}, 1395}~(2024).

\bibitem{kottmann2021tequila}
Jakob~S Kottmann, Sumner Alperin-Lea, Teresa Tamayo-Mendoza, Alba Cervera-Lierta, Cyrille Lavigne, Tzu-Ching Yen, Vladyslav Verteletskyi, Philipp Schleich, Abhinav Anand, Matthias Degroote, et~al.
\newblock ``Tequila: A platform for rapid development of quantum algorithms''.
\newblock \href{https://dx.doi.org/10.1088/2058-9565/abe567}{Quantum Science and Technology {\bf 6}, 024009}~(2021).

\bibitem{suzuki2021qulacs}
Yasunari Suzuki, Yoshiaki Kawase, Yuya Masumura, Yuria Hiraga, Masahiro Nakadai, Jiabao Chen, Ken~M Nakanishi, Kosuke Mitarai, Ryosuke Imai, Shiro Tamiya, et~al.
\newblock ``Qulacs: a fast and versatile quantum circuit simulator for research purpose''.
\newblock Quantum {\bf 5}, 559~(2021).

\bibitem{doi:10.1021/acs.jctc.2c00574}
Nick~S. Blunt, Joan Camps, Ophelia Crawford, Róbert Izsák, Sebastian Leontica, Arjun Mirani, Alexandra~E. Moylett, Sam~A. Scivier, Christoph Sünderhauf, Patrick Schopf, Jacob~M. Taylor, and Nicole Holzmann.
\newblock ``Perspective on the current state-of-the-art of quantum computing for drug discovery applications''.
\newblock \href{https://dx.doi.org/10.1021/acs.jctc.2c00574}{Journal of Chemical Theory and Computation {\bf 18}, 7001--7023}~(2022).

\bibitem{8585034}
Y.~Cao, J.~Romero, and A.~Aspuru-Guzik.
\newblock ``Potential of quantum computing for drug discovery''.
\newblock \href{https://dx.doi.org/10.1147/JRD.2018.2888987}{IBM Journal of Research and Development {\bf 62}, 6:1--6:20}~(2018).

\bibitem{von_Burg_2021}
Vera von Burg, Guang~Hao Low, Thomas Häner, Damian~S. Steiger, Markus Reiher, Martin Roetteler, and Matthias Troyer.
\newblock ``Quantum computing enhanced computational catalysis''.
\newblock Physical Review Research{\bf 3}~(2021).
\newblock  url:~\url{https://10.1103/physrevresearch.3.033055}.

\bibitem{Lordi_Nichol_2021}
Vincenzo Lordi and John~M. Nichol.
\newblock ``Advances and opportunities in materials science for scalable quantum computing''.
\newblock \href{https://dx.doi.org/10.1557/s43577-021-00133-0}{MRS Bulletin {\bf 46}, 589–595}~(2021).

\bibitem{zimborás2025mythsquantumcomputationfault}
Zoltán Zimborás, Bálint Koczor, Zoë Holmes, Elsi-Mari Borrelli, András Gilyén, Hsin-Yuan Huang, Zhenyu Cai, Antonio Acín, Leandro Aolita, Leonardo Banchi, Fernando G. S.~L. Brandão, Daniel Cavalcanti, Toby Cubitt, Sergey~N. Filippov, Guillermo García-Pérez, John Goold, Orsolya Kálmán, Elica Kyoseva, Matteo A.~C. Rossi, Boris Sokolov, Ivano Tavernelli, and Sabrina Maniscalco.
\newblock ``Myths around quantum computation before full fault tolerance: What no-go theorems rule out and what they don't''~(2025).
\newblock  \href{http://arxiv.org/abs/2501.05694}{arXiv:2501.05694}.

\bibitem{Bravyi2002}
Sergey~B. Bravyi and Alexei~Yu. Kitaev.
\newblock ``Fermionic quantum computation''.
\newblock \href{https://dx.doi.org/10.1006/aphy.2002.6254}{Annals of Physics {\bf 298}, 210--226}~(2002).

\bibitem{PRXQuantum.5.030333}
Teodor Parella-Dilm\'e, Korbinian Kottmann, Leonardo Zambrano, Luke Mortimer, Jakob~S. Kottmann, and Antonio Ac\'{\i}n.
\newblock ``Reducing entanglement with physically inspired fermion-to-qubit mappings''.
\newblock \href{https://dx.doi.org/10.1103/PRXQuantum.5.030333}{PRX Quantum {\bf 5}, 030333}~(2024).

\bibitem{algaba2025fermiontoqubitencodingsarbitrarycode}
Manuel~G. Algaba, Miha Papič, Inés de~Vega, Alessio Calzona, and Fedor Šimkovic IV.
\newblock ``Fermion-to-qubit encodings with arbitrary code distance''~(2025).
\newblock  \href{http://arxiv.org/abs/2505.02916}{arXiv:2505.02916}.

\bibitem{Vlasov_2022}
Alexander~Yurievich Vlasov.
\newblock ``Clifford algebras, spin groups and qubit trees''.
\newblock \href{https://dx.doi.org/10.12743/quanta.v11i1.199}{Quanta {\bf 11}, 97–114}~(2022).

\bibitem{miller2022bonsai}
Aaron Miller, Zolt{\'a}n Zimbor{\'a}s, Stefan Knecht, Sabrina Maniscalco, and Guillermo Garc{\'\i}a-P{\'e}rez.
\newblock ``Bonsai algorithm: Grow your own fermion-to-qubit mappings''.
\newblock \href{https://dx.doi.org/10.1103/prxquantum.4.030314}{PRX Quantum {\bf 4}, 030314}~(2023).

\bibitem{miller2024treespilationarchitecturestateoptimisedfermiontoqubit}
Aaron Miller, Adam Glos, and Zoltán Zimborás.
\newblock ``Treespilation: Architecture- and state-optimised fermion-to-qubit mappings''~(2024).
\newblock  \href{http://arxiv.org/abs/2403.03992}{arXiv:2403.03992}.

\bibitem{chiew2025optimalfermionqubitmappingsquadratic}
Mitchell Chiew, Cameron Ibrahim, Ilya Safro, and Sergii Strelchuk.
\newblock ``Optimal fermion-qubit mappings via quadratic assignment''~(2025).
\newblock  \href{http://arxiv.org/abs/2504.21636}{arXiv:2504.21636}.

\bibitem{chiew2024ternarytreetransformationsequivalent}
Mitchell Chiew, Brent Harrison, and Sergii Strelchuk.
\newblock ``Ternary tree transformations are equivalent to linear encodings of the fock basis''~(2024).
\newblock  \href{http://arxiv.org/abs/2412.07578}{arXiv:2412.07578}.

\bibitem{harrison2024sierpinskitrianglefermiontoqubittransform}
Brent Harrison, Mitchell Chiew, Jason Necaise, Andrew Projansky, Sergii Strelchuk, and James~D. Whitfield.
\newblock ``A sierpinski triangle fermion-to-qubit transform''~(2024).
\newblock  \href{http://arxiv.org/abs/2409.04348}{arXiv:2409.04348}.

\bibitem{jordan1993paulische}
Pascual Jordan and Eugene Wigner.
\newblock ``{\"U}ber das paulische {\"a}quivalenzverbot''.
\newblock Zeitschrift f{\"u}r Physik {\bf 47}, 631~(1928).

\bibitem{Coester_Kümmel_1960b}
F.~Coester and H.~Kümmel.
\newblock ``Short-range correlations in nuclear wave functions''.
\newblock \href{https://dx.doi.org/10.1016/0029-5582(60)90140-1}{Nuclear Physics {\bf 17}, 477–485}~(1960).

\bibitem{Lee_Huggins_Head-Gordon_Whaley_2018}
Joonho Lee, William~J. Huggins, Martin Head-Gordon, and K.~Birgitta Whaley.
\newblock ``Generalized unitary coupled cluster wave functions for quantum computation''.
\newblock \href{https://dx.doi.org/10.1021/acs.jctc.8b01004}{Journal of Chemical Theory and Computation {\bf 15}, 311–324}~(2018).

\bibitem{Anand_Schleich_Alperin-Lea_Jensen_Sim_Díaz-Tinoco_Kottmann_Degroote_Izmaylov_Aspuru-Guzik_2022}
Abhinav Anand, Philipp Schleich, Sumner Alperin-Lea, Phillip~W. Jensen, Sukin Sim, Manuel Díaz-Tinoco, Jakob~S. Kottmann, Matthias Degroote, Artur~F. Izmaylov, and Alán Aspuru-Guzik.
\newblock ``A quantum computing view on unitary coupled cluster theory''.
\newblock \href{https://dx.doi.org/10.1039/d1cs00932j}{Chemical Society Reviews {\bf 51}, 1659–1684}~(2022).

\bibitem{kottmann2023molecular}
Jakob~S Kottmann.
\newblock ``Molecular quantum circuit design: A graph-based approach''.
\newblock \href{https://dx.doi.org/10.22331/q-2023-08-03-1073}{Quantum {\bf 7}, 1073}~(2023).

\bibitem{doi:10.1137/0916069}
Richard~H. Byrd, Peihuang Lu, Jorge Nocedal, and Ciyou Zhu.
\newblock ``A limited memory algorithm for bound constrained optimization''.
\newblock \href{https://dx.doi.org/10.1137/0916069}{SIAM Journal on Scientific Computing {\bf 16}, 1190--1208}~(1995).
\newblock  \href{http://arxiv.org/abs/https://doi.org/10.1137/0916069}{arXiv:https://doi.org/10.1137/0916069}.

\bibitem{hibatallah2023frameworkdemonstratingpracticalquantum}
Mohamed Hibat-Allah, Marta Mauri, Juan Carrasquilla, and Alejandro Perdomo-Ortiz.
\newblock ``A framework for demonstrating practical quantum advantage: Racing quantum against classical generative models''~(2023).
\newblock  \href{http://arxiv.org/abs/2303.15626}{arXiv:2303.15626}.

\bibitem{Grant_2019}
Edward Grant, Leonard Wossnig, Mateusz Ostaszewski, and Marcello Benedetti.
\newblock ``An initialization strategy for addressing barren plateaus in parametrized quantum circuits''.
\newblock \href{https://dx.doi.org/10.22331/q-2019-12-09-214}{Quantum {\bf 3}, 214}~(2019).

\bibitem{Tranter_2019}
Andrew Tranter, Peter~J. Love, Florian Mintert, Nathan Wiebe, and Peter~V. Coveney.
\newblock ``Ordering of trotterization: Impact on errors in quantum simulation of electronic structure''.
\newblock \href{https://dx.doi.org/10.3390/e21121218}{Entropy {\bf 21}, 1218}~(2019).

\bibitem{PhysRevA.102.062612}
Yordan~S. Yordanov, David R.~M. Arvidsson-Shukur, and Crispin H.~W. Barnes.
\newblock ``Efficient quantum circuits for quantum computational chemistry''.
\newblock \href{https://dx.doi.org/10.1103/PhysRevA.102.062612}{Phys. Rev. A {\bf 102}, 062612}~(2020).

\end{thebibliography}

\onecolumn\newpage
\appendix

\section{Implementation Details}
\label{AppendixB}

\subsection{Entangling Gates Implementation}


\begin{figure*}[h]
    \centering
    \[
    \vcenter{\hbox{
    \Qcircuit @C=1em @R=1.5em {
        & \multigate{1}{\text{EG}} & \qw \\
        & \ghost{\text{EG}}        & \qw
    }
    }}
    \quad
    \vcenter{\hbox{$=$}}
    \quad
    \vcenter{\hbox{
    \Qcircuit @C=1em @R=1.5em {
        & \gate{R_y(\theta_1)} & \ctrl{1}               & \qw \\
        & \gate{R_y(\theta_2)} & \gate{R_y(\theta_3)} \qwx & \qw
    }
    }}
    \]
    \caption{Entangling gate (EG) from the CRy-HEA ansatz used in the simulation of spin systems. Decomposition into single-qubit rotations \(R_y(\theta_1)\) and \(R_y(\theta_2)\), followed by a controlled-\(R_y(\theta_3)\) gate.}
    \label{fig:block_decomposition}
\end{figure*}
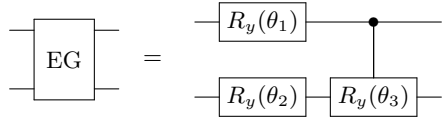

\begin{figure}[h!]
    \centering
    \begin{tikzpicture}
        \node[anchor=north west, xshift=-4em, yshift=16em]  {
            \includegraphics[width=2.2cm]{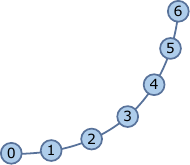}
        };
        \node[anchor=north west, xshift=11em, yshift=15.5em]  {
            \includegraphics[width=2.2cm]{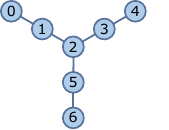}
        };
        \node[anchor=north west, xshift=29em, yshift=15em]  {
            \includegraphics[width=2.2cm]{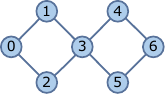}
        };
        
        \node[inner sep=0pt] (circuit) {
            $\vcenter{\hbox{
            \Qcircuit @C=1em @R=1.5em {
        & \multigate{1}{\text{EG}}     & \qw         & \qw \\
        & \ghost{\text{EG}}            &  \multigate{1}{\text{EG}}                      & \qw \\
        & \multigate{1}{\text{EG}}     & \ghost{\text{EG}}                      & \qw \\
        & \ghost{\text{EG}}            &  \multigate{1}{\text{EG}}                     & \qw \\
        & \multigate{1}{\text{EG}}     & \ghost{\text{EG}}                      & \qw \\
        & \ghost{\text{EG}}            &  \multigate{1}{\text{EG}}                      & \qw \\
        & \qw                          & \ghost{\text{EG}}                      & \qw
    }
            }}$
        };

        \node[inner sep=0pt,xshift=14em, yshift=0em] (circuit) {
            $\vcenter{\hbox{
            \Qcircuit @C=1em @R=1.5em {
    & \multigate{1}{\text{EG}}             & \qw                              & \qw                      & \qw \\
    & \ghost{\text{EG}}                    & \multigate{1}{\text{EG}}         & \qw         & \qw \\
    & \multigate{1}{\text{EG}}             & \ghost{\text{EG}}                & \gate{\text{EG}}                      & \qw \\
    & \ghost{\text{EG}}                    & \multigate{1}{\text{EG}}         & \qw& \qw\\
    & \qw                                  & \ghost{\text{EG}}                & \qw                      & \qw \\
    & \multigate{1}{\text{EG}}             & \qw                              & \gate{\text{EG}} \qwx[-3]                      & \qw \\
    & \ghost{\text{EG}}                    & \qw                              & \qw                      & \qw
}
            }}$
        };

        \node[inner sep=0pt,xshift=32em, yshift=0em] (circuit) {
            $\vcenter{\hbox{
            \Qcircuit @C=1em @R=1.5em {
    & \multigate{1}{\text{EG}}             & \gate{\text{EG}}                 & \qw                      & \qw                      & \qw \\
    & \ghost{\text{EG}}                    & \qw                              & \gate{\text{EG}}         & \qw                      & \qw \\
    & \multigate{1}{\text{EG}}             & \gate{\text{EG}} \qwx[-2]        & \qw                      & \qw                      & \qw \\
    & \ghost{\text{EG}}                    & \multigate{1}{\text{EG}}         & \gate{\text{EG}} \qwx[-2]& \gate{\text{EG}}                      & \qw\\
    & \gate{\text{EG}}                     & \ghost{\text{EG}}                & \qw                      & \qw                      & \qw \\
    & \qw                                  & \multigate{1}{\text{EG}}         & \qw                      & \gate{\text{EG}} \qwx[-2]                      & \qw \\
    & \gate{\text{EG}} \qwx[-2]            & \ghost{\text{EG}}                & \qw                      & \qw                      & \qw
}
            }}$
        };
        
    \quad
    $\vcenter{\hbox{\hspace{5pt}}}$
    \quad

        
    \end{tikzpicture}

    \caption{Entangling layers in the CRy-HEA ansatz on the exemplified 7-qubit connectivities for the simulation of spin systems. Each EG gate has its own different parameters.}
    \label{fig:block_decomposition}
\end{figure}

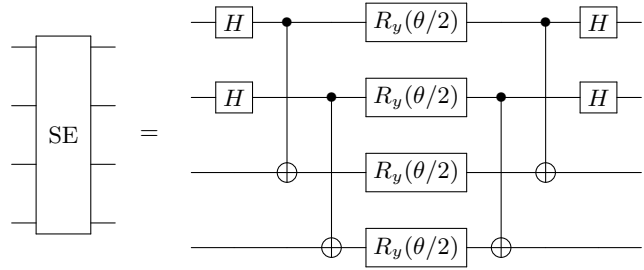
\begin{figure*}[h]
    \centering
    \[
    \vcenter{\hbox{
    \Qcircuit @C=1em @R=1.5em {
        & \multigate{3}{\text{SE}} & \qw \\
        & \ghost{\text{SE}}        & \qw \\
        & \ghost{\text{SE}}        & \qw \\
        & \ghost{\text{SE}}        & \qw
    }
    }}
    \quad
    \vcenter{\hbox{$=$}}
    \quad
    \vcenter{\hbox{
    \Qcircuit @C=1em @R=1.5em {
        & \gate{H}           & \ctrl{2} & \qw       & \gate{R_y(\theta/2)} & \qw& \ctrl{2}& \gate{H}& \qw \\
        & \gate{H}           & \qw& \ctrl{2}        & \gate{R_y(\theta/2)} & \ctrl{2} & \qw& \gate{H}& \qw \\
        & \qw               & \targ     & \qw      & \gate{R_y(\theta/2)} & \qw& \targ& \qw& \qw \\
        & \qw               & \qw& \targ           & \gate{R_y(\theta/2)} & \targ     & \qw& \qw& \qw
    }
    }}
    \]
    \caption{Decomposition of single excitation gate (SE) used in excitation-based circuits for electronic structure simulation, acting on four qubits and parametrized by a single parameter $\theta$.}
    \label{fig:single_excitation_decomp}
\end{figure*}

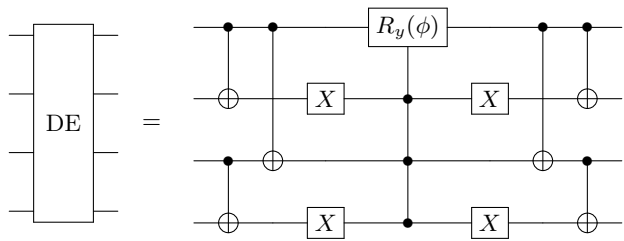
\begin{figure*}[h]
    \centering
    \[
    \vcenter{\hbox{
    \Qcircuit @C=1em @R=1.5em {
        & \multigate{3}{\text{DE}} & \qw \\
        & \ghost{\text{DE}}        & \qw \\
        & \ghost{\text{DE}}        & \qw \\
        & \ghost{\text{DE}}        & \qw
    }
    }}
    \quad
    \vcenter{\hbox{$=$}}
    \quad
    \vcenter{\hbox{
\Qcircuit @C=1em @R=1.5em {
    & \ctrl{1} & \ctrl{2} & \qw      & \gate{R_y(\phi)}     & \qw      & \ctrl{2} & \ctrl{1} & \qw \\
    & \targ    & \qw      & \gate{X} & \control \qw & \gate{X} & \qw      & \targ    & \qw \\
    & \ctrl{1}      & \targ    & \qw      & \control \qw & \qw      & \targ    & \ctrl{1}      & \qw \\
    & \targ      & \qw      & \gate{X} & \control \qwx[-3] \qw & \gate{X} & \qw      & \targ      & \qw
}
}}
\]

    \caption{Decomposition of a double excitation gate (DE) used in excitation-based circuits for electronic structure simulation, acting on four qubits and parametrized by a single parameter $\phi$.}
    \label{fig:double_excitation_decomp}
\end{figure*}

\begin{figure*}[h]
    \centering
    \[
    \vcenter{\hbox{
    \Qcircuit @C=1em @R=1.5em {
        & \multigate{3}{\text{EG'}} & \qw \\
        & \ghost{\text{EG'}}        & \qw \\
        & \ghost{\text{EG'}}        & \qw \\
        & \ghost{\text{EG'}}        & \qw
    }
    }}
    \quad
    \vcenter{\hbox{$=$}}
    \quad
    \vcenter{\hbox{
    \Qcircuit @C=1em @R=1.5em {
        & \multigate{3}{\text{DE}} & \qw & \multigate{3}{\text{SE}}& \qw\\
        & \ghost{\text{DE}}        & \qw & \ghost{\text{SE}}& \qw\\
        & \ghost{\text{DE}}        & \qw & \ghost{\text{SE}}& \qw\\
        & \ghost{\text{DE}}        & \qw& \ghost{\text{SE}}& \qw
    }
    }}
\]
    \caption{Decomposition of an entangling gate (EG') used in excitation-based circuits for electronic structure simulation. It acts on four qubits, and consists in a double excitation followed by a single excitation, parametrized by two parameters $\phi$ (DE) and $\theta$ (SE).}
    \label{fig:EG_excitation_decomp}
\end{figure*}
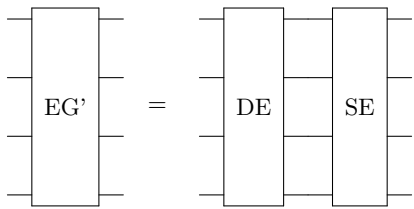

\begin{figure}[h!]
    \centering
    \begin{tikzpicture}
        \node[anchor=north west, xshift=-25em, yshift=5em]  {
            \includegraphics[width=7cm]{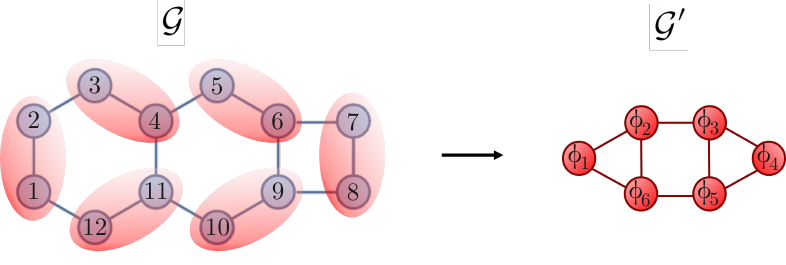}
        };

    \node[inner sep=0pt,xshift=8em, yshift=0em] (circuit) {
            $\vcenter{\hbox{
            \Qcircuit @C=1em @R=1.5em {
    \lstick{\ket{\phi_1,\alpha}}& \multigate{3}{\text{EG'}}  & \multigate{1}{\text{EG'}}        & \qw                      & \qw                              & \qw & \qw \\
    \lstick{\ket{\phi_1,\beta}}& \ghost{\text{EG'}}         & \ghost{\text{EG'}}               & \qw                      & \qw                              & \qw & \qw \\
    \lstick{\ket{\phi_2,\alpha}}& \ghost{\text{EG'}}         & \qw\                            & \multigate{3}{\text{EG'}} & \multigate{1}{\text{EG'}}         & \qw & \qw \\
    \lstick{\ket{\phi_2,\beta}}& \ghost{\text{EG'}}         & \qw                             & \ghost{\text{EG'}}        & \ghost{\text{EG'}}                & \qw & \qw \\
    \lstick{\ket{\phi_3,\alpha}}& \multigate{3}{\text{EG'}}  & \qw                             & \ghost{\text{EG'}}        & \qw                              & \multigate{1}{\text{EG'}} & \qw \\
    \lstick{\ket{\phi_3,\beta}}& \ghost{\text{EG'}}         & \qw                             & \ghost{\text{EG'}}        & \qw                              & \ghost{\text{EG'}} & \qw \\
    \lstick{\ket{\phi_4,\alpha}}& \ghost{\text{EG'}}         & \qw                             & \multigate{3}{\text{EG'}} & \qw                              & \qw & \qw \\
    \lstick{\ket{\phi_4,\beta}}& \ghost{\text{EG'}}         & \qw                             & \ghost{\text{EG'}}        & \qw                              & \qw & \qw \\
    \lstick{\ket{\phi_5,\alpha}}& \multigate{3}{\text{EG'}}  & \qw                             & \ghost{\text{EG'}}        & \qw                              & \multigate{1}{\text{EG'}}\qwx[-3] & \qw \\
    \lstick{\ket{\phi_5,\beta}}& \ghost{\text{EG'}}         & \qw                             & \ghost{\text{EG'}}        & \qw                              & \ghost{\text{EG'}} & \qw \\ 
    \lstick{\ket{\phi_6,\alpha}}& \ghost{\text{EG'}}         & \multigate{1}{\text{EG'}}\qwx[-9]& \qw                      & \multigate{1}{\text{EG'}}\qwx[-7] & \qw & \qw \\
    \lstick{\ket{\phi_6,\beta}}& \ghost{\text{EG'}}         & \ghost{\text{EG'}}               & \qw                      & \ghost{\text{EG'}}                & \qw & \qw 
    }
            }}$
        };
        
    \quad
    $\vcenter{\hbox{\hspace{5pt}}}$
    \quad

        
    \end{tikzpicture}

    \caption{Entangling layer in the excitation-based ansatz used for electronic structure simulation in the exemplified connectivity. The qubit connectivity $\mathcal{G}$ (blue) is coarsened pairwise to get a connectivity in the molecular orbital level $\mathcal{G}'$ (red). Each EG gate has its own two different parameters. The depth in EG gates is equal to the maximum connectivity of a single node in $\mathcal{G}'$ (3 in the exemple).}
    \label{fig:block_decomposition}
\end{figure}

\FloatBarrier
\subsection{Swap Networks Implementation
}

\begin{figure*}[h]
    \centering
    \[
    \vcenter{\hbox{
    \Qcircuit @C=1em @R=1em {
        & \multigate{1}{\text{SWAP}} & \qw \\
        & \ghost{\text{SWAP}} & \qw
    }
    }}
    \vcenter{\hbox{\hspace{15pt}$=$\hspace{15pt}}}
    \vcenter{\hbox{
    \Qcircuit @C=1em @R=1em {
        & \targ     & \ctrl{1} & \targ     & \qw \\
        & \ctrl{-1} & \targ    & \ctrl{-1} & \qw
    }
    }}
    \vcenter{\hbox{\hspace{15pt}$=$\hspace{15pt}}}
    \vcenter{\hbox{
    \Qcircuit @C=1em @R=1em {
        & \qw & \ctrl{1} & \gate{H} & \ctrl{1} & \gate{H} & \ctrl{1}& \qw& \qw \\
        & \gate{H} & \ctrl{-1} & \gate{H} & \ctrl{-1} & \gate{H} & \ctrl{-1}& \gate{H}& \qw
    }
    }}
    \]
    \caption{SWAP gate and its equivalent decompositions into CNOT gates, and CZ + Hadamard gates.}
    \label{fig:swap_decompositions}
\end{figure*}
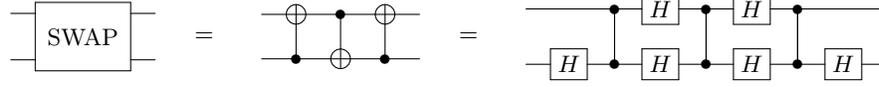

\begin{figure*}[h]
    \centering
    \[
    \vcenter{\hbox{
        \Qcircuit @C=1em @R=2em {
            & \qswap        & \qw \\
            & \qswap \qwx   & \qw
        }
    }}
    \quad
    \vcenter{\hbox{$=$}}
    \quad
    \vcenter{\hbox{
        \Qcircuit @C=1em @R=1.2em {
            & \targ     & \ctrl{1} & \targ     & \qw \\
            & \ctrl{-1} & \targ    & \ctrl{-1} & \qw
        }
    }}
    \quad
    \vcenter{\hbox{$=$}}
    \quad
    \vcenter{\hbox{
        \Qcircuit @C=1em @R=1em {
            & \gate{H} & \ctrl{1}    & \gate{H} & \ctrl{1}   & \gate{H} & \ctrl{1}    & \gate{H} & \qw \\
            & \qw      & \ctrl{-1}   & \gate{H} & \ctrl{-1}   & \gate{H} & \ctrl{-1}   & \qw      & \qw
        }
    }}
    \]
    \caption{SWAP gate decomposition into three CNOT gates, or 3 CZ + 6 Hadamard gates.}
    \label{fig:swapcnot_decompositions}
\end{figure*}

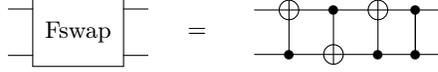
\begin{figure*}[h]
    \centering
    \[
    \vcenter{\hbox{
    \Qcircuit @C=1em @R=1em {
        & \multigate{1}{\text{Fswap}} & \qw \\
         & \ghost{\text{Fswap}} & \qw
    }
    }}
    \vcenter{\hbox{\hspace{15pt}$=$\hspace{15pt}}}
    \vcenter{\hbox{
    \Qcircuit @C=1em @R=1em {
         & \targ     & \ctrl{1} & \targ     & \ctrl{1} & \qw \\
          & \ctrl{-1} & \targ    & \ctrl{-1} & \ctrl{-1}& \qw
    }
    }}
    \]
    \caption{Fermionic Swap gate (Fswap) and its standard decomposition.}
    \label{fig:fswap}
\end{figure*}

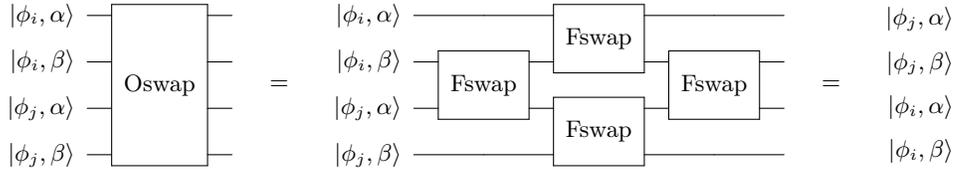
\begin{figure*}[h]
    \[
\vcenter{\hbox{
\Qcircuit @C=1em @R=1em {
    \lstick{\ket{\phi_i,\alpha}} & \multigate{3}{\text{Oswap}} & \qw \\
    \lstick{\ket{\phi_i,\beta}}& \ghost{\text{Oswap}}        & \qw \\
    \lstick{\ket{\phi_j,\alpha}}& \ghost{\text{Oswap}}        & \qw \\
    \lstick{\ket{\phi_j,\beta}}& \ghost{\text{Oswap}}        & \qw
}
}}
\quad
\vcenter{\hbox{\hspace{5pt}$=$\hspace{35pt}}}
\quad
\vcenter{\hbox{
\Qcircuit @C=1em @R=1em {
   \lstick{\ket{\phi_i,\alpha}} & \qw                           & \multigate{1}{\text{Fswap}} & \qw                         & \qw \\
    \lstick{\ket{\phi_i,\beta}}& \multigate{1}{\text{Fswap}}   & \ghost{\text{Fswap}}        & \multigate{1}{\text{Fswap}} & \qw \\
    \lstick{\ket{\phi_j,\alpha}}& \ghost{\text{Fswap}}    & \multigate{1}{\text{Fswap}} & \ghost{\text{Fswap}}        &  \qw \\
    \lstick{\ket{\phi_j,\beta}}& \qw                           & \ghost{\text{Fswap}}        & \qw                         & \qw
}
}}
\quad
\vcenter{\hbox{\hspace{5pt}$=$\hspace{35pt}}}
\quad
\vcenter{\hbox{
\Qcircuit @C=1em @R=1.8em {
   \lstick{\ket{\phi_j,\alpha}}  \\
    \lstick{\ket{\phi_j,\beta}} \\
    \lstick{\ket{\phi_i,\alpha}} \\
    \lstick{\ket{\phi_i,\beta}}
}
}}
\]
    \caption{Decomposition of the Orbital Swap (Oswap) gate using Fswap gates. The Oswap gate exchanges the orbitals $\phi_i$ and $\phi_j$ in qubit representation.}
    \label{fig:oswap}
\end{figure*}


\FloatBarrier
\newpage
\subsection{Electronic Structure Initialization}

\begin{table*}[h]
    \centering
    \begin{tabular}{ |p{2cm}|p{2cm}|p{2.5cm}|p{4cm}|  }
 \hline
 Molecule & Basis  & $E^{corr.}(kcal/mol)$ & Cartesian Geometry (Å)\\
 \hline 
 \hline
 p-benzyne   & \textit{cc-pVDZ}  & 76.7520 & C -0.7396 -1.1953 0.0000 \\
   &  &  & C 0.7396 -1.1953 0.0000  \\
   &  &  & C 1.3620 0.0000 0.0000 \\
   &  &  & C 0.7396 1.1953 0.0000 \\
   &  &  & C -0.7396 1.1953 0.0000  \\
   &  &  &  C -1.3620 0.0000 0.0000 \\
   &  &  &  H 1.1999 -2.1824 0.0000 \\
   &  &  &  H -1.1999 2.1824 0.0000 \\
   &  &  &  H 1.1999 2.1824 0.0000 \\
   &  &  & H -1.1999 -2.1824 0.0000 \\
   \hline

\end{tabular}
    
    \caption{Initialization details for the $\pi$ active space system of p-benzyne birradical. }
    \label{tab:geometry}
\end{table*}

\begin{figure*}[h]                       
\centering
\includegraphics[width=.5
\columnwidth, angle=0]{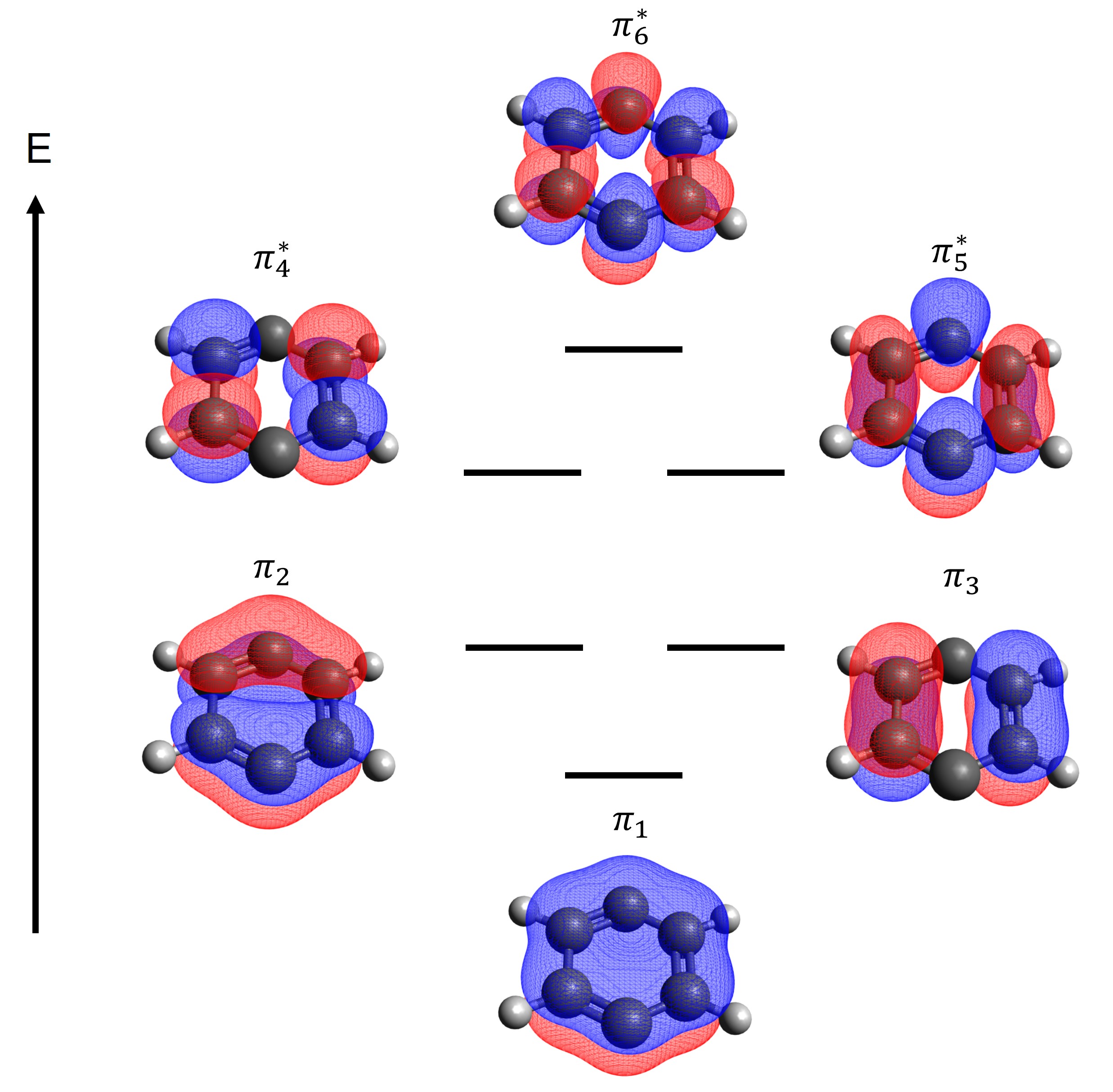}
\caption{Active space of $\pi$ molecular orbitals of p-benzyne, in the basis of canonical orbitals.}
\label{fig:Benzene_Orbitals}
\end{figure*}

\newpage

\FloatBarrier
\subsection{Graph examples}

\begin{figure*}[h]                       
\centering
\includegraphics[width=1
\columnwidth, angle=0]{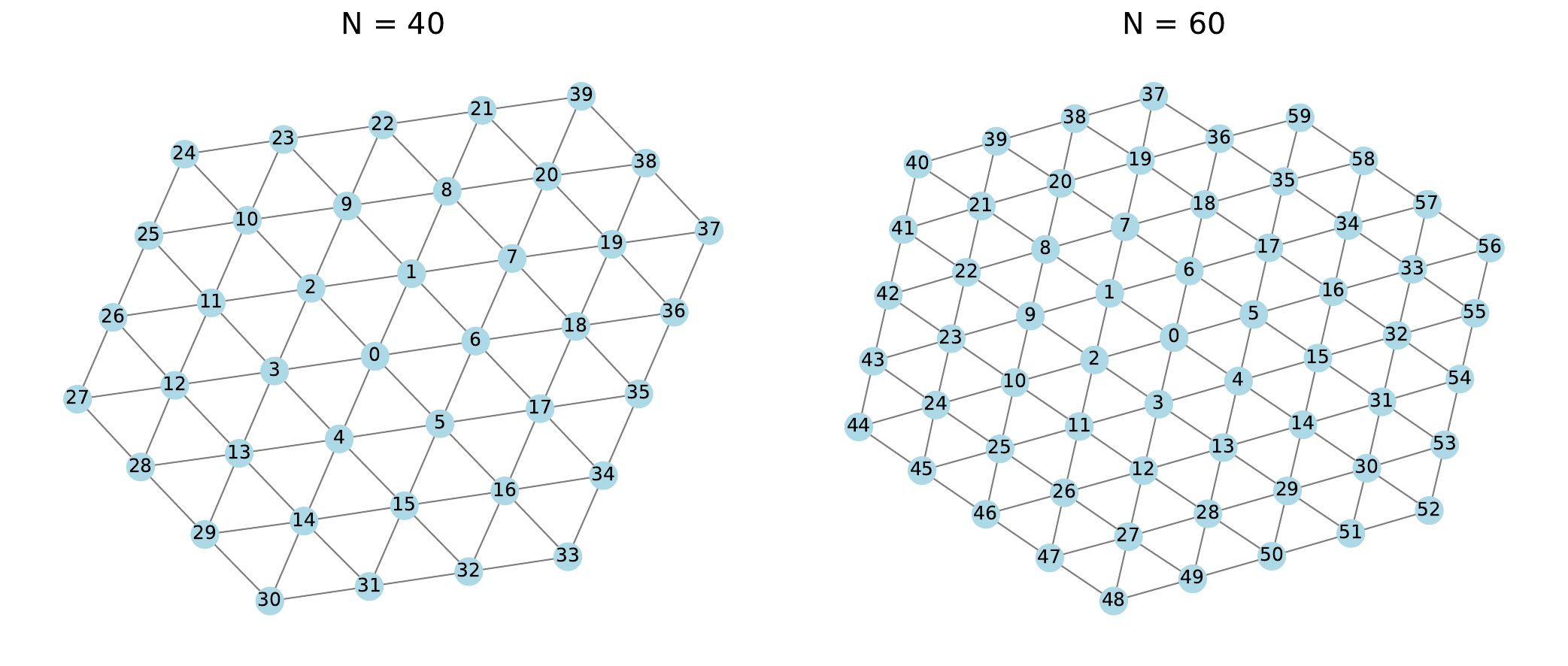}
\caption{Finite graph examples from the graph family in Fig.\ref{fig:SquareGridCoarsening}, generated with the function $\mathit{create\_hex\_triangular\_grid}$.}
\label{fig:GraphsExamples}
\end{figure*}

\FloatBarrier
\subsection{Optimization Parameters}

\begin{table*}[h]
    \centering
    
    \begin{tabular}{ | p{2cm}|p{1cm}|p{1cm}|p{1cm}|p{3cm}| p{3cm}| }
 \hline
 System & $T_0$ & $k$  & $\alpha$ & Annealing Steps & Annealings\\
 \hline 
 \hline
 spin glasses & 10 & 2  & 2 & 1000 & 10\\
  \hline
 p-benzyne  & 10 & 2  & 2 & 1000 & 10\\
 \hline
 Fig.\ref{fig:Nq_vs_Iterations}  & 1 & 1  & 1 & 1000 & 10,50,100,500,1000\\
   \hline
 Fig.\ref{fig:SabreComparison}  & 1 & 1  & 1 & 1000 & 100\\
    \hline
 Fig.\ref{fig:Alphadependance}  & 1 & 1  & 0.1 - 3 & 1000 & 100\\

   \hline

\end{tabular}
    
    \caption{Parameters used in the algorithm optimizations. }
    \label{tab:parameters}
\end{table*}

\begin{table*}
    
\end{table*}
\newpage
\FloatBarrier
\subsection{Exponent Dependence}

\begin{figure*}[h]                       
\centering
\includegraphics[width=.65
\columnwidth, angle=0]{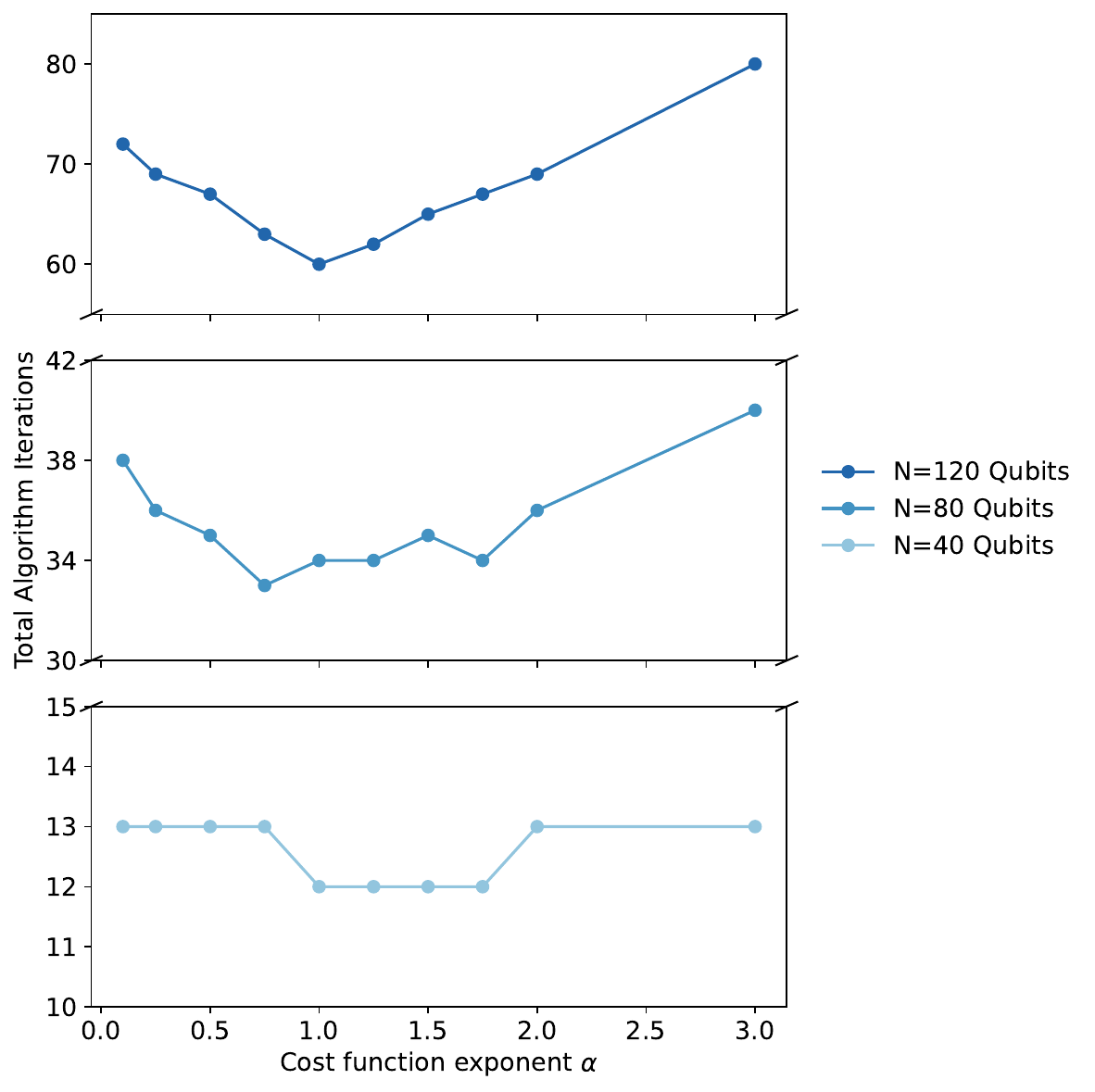}
\caption{Analysis of the total algorithm iterations dependence on the exponent $\alpha$ in Eq.~\ref{CostFunction}. Optimization results for three graph instances from the hexagonal family shown in Figs.~\ref{fig:SquareGridCoarsening} and \ref{fig:GraphsExamples}. The total annealing time is the same for all simulations. The dependence on the exponent becomes more pronounced for larger systems, as seen for $N=12$. Since $\alpha=1$ consistently yields among the best performance in our simulations, we adopt this value for the large graph scaling studies, as shown in Tab.\ref{tab:parameters}.
}
\label{fig:Alphadependance}
\end{figure*}

\newpage

\section{Approximate Implementation of Fermionic Excitations}
\label{AppendixA}
In the UCC theory \cite{Coester_Kümmel_1960b}, a reference state is parametrized in the form:

\begin{equation}
    |\psi_{UCC}\rangle = e^{T-T^{\dagger}} |\psi_{ref.}\rangle
\end{equation}
with the cluster operator:
\begin{equation}
    T = \sum_{i,j} t_{i}^{j} \hat{a}_i^\dagger \hat{a}_j +\sum_{i,j} t_{i,j}^{k,l} \hat{a}_i^\dagger \hat{a}_j^\dagger \hat{a}_k \hat{a}_l+...
\end{equation}
with $t_{i}^{j}$, $t_{i,j}^{k,l}$,... coefficients. Typically, the UCC can be restricted to single and double excitations (explicit terms in the previous equation), yielding the UCC of singles and doubles (UCCSD). Higher order excitations can be added in the description to yield higher order UCC forms, such as triples, quadruples... In the following, we consider the UCCSD form. The operation $ e^{T-T^{\dagger}}$ is typically implemented via Suzuki–Trotter decomposition as $q$ Trotter steps, in the form of individual single and double excitations:
\begin{equation}
    |\psi_{UCCSD}\rangle \simeq \bigg[ \bigg ( \prod_{i < j}
    e^{\frac{t_i^j}{q} (\hat{a}_i^\dagger \hat{a}_j - \hat{a}_i \hat{a}_j^\dagger)}  \bigg ) 
    \bigg(
    \prod_{i < j<n< m}
    e^{\frac{t_{i,j}^{n,m}}{q} (\hat{a}_{i}^\dagger\hat{a}_j^\dagger  \hat{a}_n \hat{a}_m -  \hat{a}_i  \hat{a}_j \hat{a}_{n}^\dagger \hat{a}_m^\dagger)}  \bigg ) \bigg]^q
    |\psi_{ref.}\rangle
\end{equation}
being an equality at $\lim q \to \infty$. In practice, however, one usually sets $q=1$ to minimize implementation resources. The standard UCCSD ansatz exhibits steep circuit depth and a rapidly growing parameter space, as the amount of excitations is $\mathcal{O}(N^4)$. Moreover, the non-commutativity between the excitations makes its ordering influence the simulation \cite{Tranter_2019}.
\\
\\
Additional improvements can be performed on the UCCSD, such as restricting excitations only between Fermionic modes associated to two spatial orbitals. Repeating such a complete circuit $k$ times, yields the $k$-unitary paired coupled cluster of generalized singles and doubles ($k$-UpCCGSD), which provides a more compact, scalable, and hardware-friendly alternative. In $k$-UpCCGSD, double excitations are restricted to the four spin-orbitals $\phi_a^{\alpha},\phi_a^{\beta},\phi_b^{\alpha},\phi_b^{\beta}$ associated to two different molecular orbitals $\phi_a,\phi_b$, with spin $\alpha$ / $\beta$. That is, the only non-zero coefficients can be $t_{\phi_a^{\alpha},\phi_a^{\beta}}^{\phi_b^{\alpha},\phi_b^{\beta}}$. Additionally, in the closed-shell version, single excitations are symmetric between spin-orbitals of equal molecular orbitals $t_{\phi_a^{\alpha}}^{\phi_b^{\alpha}} = t_{\phi_a^{\beta}}^{\phi_b^{\beta}}$.
\\
\\
 We now consider the spin-orbital ordering $\{\phi_1^{\alpha},\phi_2^{\alpha}, ... , \phi_N^{\alpha},\phi_1^{\beta},\phi_2^{\beta}, ... , \phi_N^{\beta}\}\equiv \{1,2,...,N,N+1,N+2,..., 2N\}$ under the JW transformation, where ladder operators are transformed as:
 
 \begin{align}
a_j^\dagger = \left( \prod_{k=0}^{j-1} Z_k \right) \cdot \frac{X_j - iY_j}{2} \hspace{20pt},\hspace{20pt}
a_j = \left( \prod_{k=0}^{j-1} Z_k \right) \cdot \frac{X_j + iY_j}{2}
\end{align}

 The implementation of single (double) excitations between adjacent molecular orbitals $\phi_i$ and $\phi_{i+1}$, result in 2 (4) qubit operations due to the cancellation of long Z Pauli strings from the ladder operators. In this direction, UpCCGSD in a ladder qubit connectivity can be implemented using 4-qubit entanglers, where single excitations between two molecular orbitals can be implemented with 4 CNOTs, while double excitations can be implemented with 13 CNOT gates \cite{PhysRevA.102.062612,kottmann2021tequila}. 
\\
\\
In our setting with an arbitrary qubit connectivity graph, we may wish to implement excitation operators between modes associated with molecular orbitals $\psi_i$ and $\psi_j$ that are adjacent on the hardware graph but non‑contiguous in the Fermionic ordering (i.e., $i<j$ and $j\neq i+1$). In this case, the exact implementation of single Fermionic excitations act on $1+j-i$ qubits (or $2(1+j-i)$ qubits in the case of double excitations), increasing significantly the resources. 
\\
\\
Drawing inspiration from qubit-excitation-based excitations \cite{Yordanov_Armaos_Barnes_Arvidsson-Shukur_2021} or qubit excitations \cite{PRXQuantum.2.020310} from ADAPT-VQE methods \cite{Grimsley_Economou_Barnes_Mayhall_2019}, we implement an approximation of the Fermionic excitations in exchange to improved hardware-friendliness of the ansatz. To do so, we truncate the parity qubit sets on the pauli strings, that is, the long Z strings part. Therefore, single excitations are implemented between arbitary orbitals $\phi_p$, $\phi_q$ as:

\begin{align}
    U^{single}_{\phi_p,\phi_q}(t_p^q) \simeq \exp\bigg[ i\frac{t_p^q}{4} \bigg( & X_pY_q-Y_pX_q+X_{N+p}Y_{N+q}-Y_{N+p}X_{N+q}
       \bigg)\bigg]
    \label{QEB2}
\end{align}

And double excitations as:

\begin{align}
    U^{double}_{\phi_j,\phi_k}(t_{p,N+p}^{  q,N+q}) \simeq \exp\bigg[ i\frac{t_{p,N+p}^{  q,N+q}}{8} \bigg( & Y_pX_{N+p}X_qX_{N+q}+X_pY_{N+p}X_qX_{N+q}-X_pX_{N+p}Y_qX_{N+q}\nonumber\\
    &-X_pX_{N+p}X_qY_{N+q}+Y_pY_{N+p}X_qY_{N+q}+Y_pY_{N+p}Y_qX_{N+q}\nonumber\\
    &-X_pY_{N+p}Y_qY_{N+q}-Y_pX_{N+p}Y_qY_{N+q}
       \bigg)\bigg]
    \label{QEB2}
\end{align}
being equivalent as in the case of two adjacent qubits in the Fermionic enumeration. That is, any single (double) excitation gate in the arbitrary connectivity setting, is implemented with 4 (13) CNOT gates.

\end{document}